\documentclass[english]{article}
\usepackage[toc,page]{appendix}
\usepackage[T1]{fontenc}
\usepackage{pgfplots}
\usepackage[utf8]{inputenc}
\usepackage{xcolor,mdframed}
\usepackage[a4paper]{geometry}
\usepackage{babel}
\usepackage{xcolor}
\usepackage{float}
\usepackage{graphicx}
\usepackage{units}
\usepackage{amsmath}
\usepackage{amsthm}
\usepackage{amssymb}
\usepackage{setspace}
\usepackage{xcolor}
\pgfplotsset{compat=1.15}
\usepackage[unicode=true,pdfusetitle,
 bookmarks=true,bookmarksnumbered=false,bookmarksopen=false,
 breaklinks=false,pdfborder={0 0 1},backref=false,colorlinks=false]
 {hyperref}

\definecolor{mycolor}{RGB}{119, 38, 222}

 \newcommand{\Tr}[1]{\mathrm{Tr}\left(#1\right)}

 \newcommand{\dd}{\mathrm{d}}

\usepackage{tikz}
\usepackage{pgfplots}
\usepgfplotslibrary{external}
\tikzset{external/system call={lualatex --shell-escape
		\tikzexternalcheckshellescape --halt-on-error --interaction=batchmode
		--jobname "\image" "\texsource"}}
\tikzset{cross/.style={cross out, draw=black, minimum size=2*(#1-\pgflinewidth), inner sep=0pt, outer sep=0pt},
	cross/.default={2.2pt}}

\definecolor{abricot}{RGB}{230, 126, 48}
\definecolor{violeteminence}{RGB}{108,48,130}

\usetikzlibrary{shapes.misc}

\makeatletter

\usepackage{feyn}
\usepackage{lmodern}
\usepackage{float}
\renewcommand\[{\begin{equation}}
\renewcommand\]{\end{equation}}
\hypersetup{pdfauthor={Name}}
\makeatother
\title{Searching for continuous phase transitions\\ in 5D SU(2) lattice gauge theory
}
\author{Adrien Florio$^1$, João M. Viana P. Lopes $^2$, José Matos $^2$, João Penedones$^3$}
\date{\small \it$^1$ Center for Nuclear Theory, Department of Physics and Astronomy,  \\ Stony Brook University, Stony Brook, New York 11794, USA.\\
$^2$Centro de Física das Universidades do Porto, University of Porto, 4169-007 Porto, Portugal\\
$^3$Field and String Laboratory, Institute of Physics, \\ \'Ecole Polytechnique F\'ed\'erale de Lausanne (EPFL), CH-1015 Lausanne, Switzerland.
}

\begin{document}

\maketitle

\begin{abstract}
We study the phase diagram of 5-dimensional $SU(2)$ Yang-Mills theory on the lattice.
We consider two extensions of the fundamental plaquette Wilson action in the search for the continuous phase transition suggested by the $4+\epsilon$ expansion.
The extensions correspond to new terms in the action: \emph{i}) a unit size plaquette in the adjoint representation or \emph{ii}) a two-unit sided square plaquette in the fundamental representation.
We use Monte Carlo to sample the first and second derivative of the
entropy near the confinement phase transition,
with lattices up to $12^{5}$.
While we exclude the presence of a second order phase transition in the parameter space we sampled for model \emph{i)}, our data is not conclusive in some regions of the parameter space of model \emph{ii)}.

\end{abstract}

\section{Introduction}

An important open problem of theoretical physics is the classification of consistent field theories. To study this problem, the space of field theories can be given a structure in the framework of the Renormalization Group (RG); RG flows now link different theories, see~\cite{Wilson:1973jj} for a review. Fixed points of the RG are special as they are the start and end-points of RG flows; they can be thought of as "signposts" \cite{Poland:2018epd} in the space of consistent field theories. As a result, an essential step towards the classification of field theories is the classification of Conformal Field Theories, which describe  RG fixed points. This classification is the main aim of the Conformal Bootstrap program, reviewed in \cite{Poland:2018epd}.

In this light, a legitimate endeavor is to look for the existence of non-trivial CFTs.
These become  rarer as one increases the spacetime dimension. In fact, there is no known interacting CFT in $D>6$.
On the other hand, the $\epsilon$ expansion suggests the existence of a non-trivial fixed point for Yang-Mills (YM) in $D=4+\epsilon$, as originally pointed out in \cite{Peskin:1980ay}.
This is a UV fixed point similar to  the non-linear $\sigma$ model in $2+\epsilon$ dimensions \cite{Wilson:1973jj,Polyakov:1975rr,Bardeen:1976zh,Arefeva:1978fj}.
Such fixed points motivate the   "asymptotic safety" program, see \cite{Eichhorn:2018yfc} for a review.
A question that has not been conclusively answered so far is whether this interacting YM fixed point exists in $D=5$ and possibly higher dimensions. An investigation using the functional renormalization group approach was performed in \cite{Gies:2003ic} and found the fixed point likely to exist in five dimensions. Another study favoring the existence of such a fixed point was conducted in \cite{Morris:2004mg, DeCesare:2021pfb}, where the $\epsilon$ expansion was computed up to order $\epsilon^4$ and $\epsilon^5$ respectively.
In addition, the paper \cite{BenettiGenolini:2019zth} proposes that such a fixed point can be found in the IR limit of an RG flow starting from a known superconformal field theory.
On the other hand, attempts to directly find and study this theory using lattice models have so-far been inconclusive \cite{Creutz:1979dw,Kawai:1992um}, see also \cite{Nishimura:1996pe,Ejiri:2000fc,Ejiri:2002ww,Farakos:2002zb,deForcrand:2010be, DelDebbio:2013rka,Knechtli:2016pph} for exploration with a compact dimension and \cite{Kanazawa:2014fla} for non-Lorentz invariant extensions.

The expectation from the $\epsilon$ expansion is that the interacting YM   fixed point in $D=5$  has only one relevant gauge invariant operator.
\footnote{Another possibility is that there are relevant  monopoles operators charged under the topological $U(1)$ symmetry associated with the conserved current $j_\mu = \epsilon_{\mu\nu\rho\alpha\beta}F^{\nu\rho}F^{\alpha\beta}$.
We thank the participants of this \href{https://youtu.be/Bh0Air1grCs}{\underline{discussion}}
in the Bootstrap 2021 workshop for bringing this scenario to our attention.
}
Therefore, the critical surface should have co-dimension one and the corresponding continuous phase transition should be   generically  found by tuning only one coupling.
However, it is well known that there is no continuous phase transition in the standard Wilson lattice gauge theory in $5D$   \cite{Creutz:1979dw}.
For this reason, we consider two modifications of the lattice action using the adjoint representation and plaquettes  $2\times 2$ (see equations \eqref{Fundamental + Adjoint Action} and \eqref{2 Loops Action}).
The spirit of our approach is very similar to the  study \cite{Kawai:1992um}, which was inconclusive due to the difficulty of distinguishing weakly first order from continuous phase transitions.
In this work, benefiting from almost 30 years of Moore's law, we revisit this issue.

\section{Actions, observables and algorithms}\label{subsec:models}

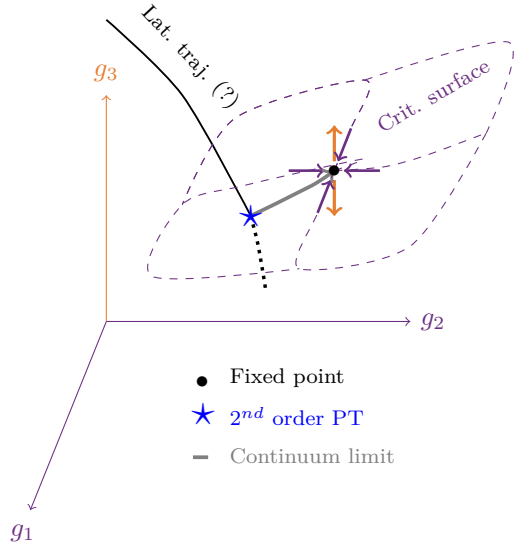
\begin{figure}
\centering
\begin{tikzpicture}
\draw[->,color=abricot] (1,2) -- (1,5);
\node[color=abricot] at (1,5.3) {$g_3$};
\draw[->, color=violeteminence] (1,2) -- (5,2);
\draw[->, color=violeteminence] (1,2) -- (0,-0.5);
\node[color=violeteminence] at (-0.1,-0.8) {$g_1$};
\node[color=violeteminence] at (5.3,2) {$g_2$};

\node (A) at (4,4){};

%relevant
\draw[color=abricot,->, line width = 1.2pt] (A) -- (4,4.6);
\draw[color=abricot,->, line width = 1.2pt] (A) -- (4,3.4);

%irrelevant
\draw[color=violeteminence,<-, line width = 1.pt] (A) -- (4.6,4);
\draw[color=violeteminence,<-, line width = 1.pt] (A) -- (3.4,4);

\draw[color=violeteminence,<-, line width = 1.pt] (A) -- (3.78,3.44);
\draw[color=violeteminence,<-, line width = 1.pt] (A) -- (4.22,4.56);

%lattice trajectory

\draw[line width=0.7pt] plot [smooth] coordinates {(1,6) (2,5) (2.9,3.4)};
\draw[dotted, line width=1.3pt] plot [smooth] coordinates {(2.9,3.4) (3., 3.0) (3.1, 2.4)};

%TO COMMENT to remove the continuum limit
\draw[color=gray, line width=1.4pt, ->] plot [smooth] coordinates {(2.9,3.4) (3.7, 3.8) (A)};

%critical surface
\draw[color=violeteminence, dashed] plot [smooth] coordinates {(A) (4.65,4.05) (5.2,4.2)  (6,4.5)};
\draw[color=violeteminence, dashed] plot [smooth] coordinates {(A) (4.35,4.15) (2.8,3.8)  (2,3.55) (1.6,2.6) (3.54,2.7) };
\draw[color=violeteminence, dashed] plot [smooth] coordinates {(A) (3.78,3.44) (3.54,2.7) (5,3)  (6,4.5)};
\draw[color=violeteminence, dashed] plot [smooth] coordinates {(A) (4.22,4.56) (4.4,5.2) (2.7,4.7) (2,3.55)};
\draw[color=violeteminence, dashed] plot [smooth] coordinates {(A) (4.22,4.56) (4.4,5.2) (2.7,4.7) (2,3.55)};
\draw[color=violeteminence, dashed] plot [smooth] coordinates {(4.4,5.2) (6.3,5.7) (6,4.5)};

% critical point

\node at (A) {$\bullet$};

%lattice 2nd order
\draw node[color=blue] at (2.9,3.4) {\LARGE$\star$};

%legend
\node[anchor=south west,color=black] at (2.05,1) {$\bullet$};
\node[anchor=south west,color=black] at (2.5,1) {\footnotesize Fixed point};

\node[anchor=south west,color=blue] at (2.,0.5) {\LARGE$\star$};
\node[anchor=south west,color=blue] at (2.5,0.5) {\footnotesize $2^{nd}$ order PT};

%TO COMMENT to remove the continuum limit
\draw[color=gray, line width = 1.4pt ] (2.14,0.2) -- (2.34,0.2);
%TO COMMENT to remove the continuum limit
\node[anchor=south west,color=gray] at (2.5,0.) {\footnotesize Continuum limit};

\node[rotate=313] at (2.1,5.5){\footnotesize Lat. traj. (?)};
\node[rotate=30, color=violeteminence] at (5.3,5){\footnotesize Crit. surface};

\end{tikzpicture}
\caption{Sketch of the fixed point scenario for 5D $SU(2)$ Yang-Mills suggested by the $\epsilon$ expansion. The orange (purple) arrows represent the repulsive (attractive) directions. The black line is the trajectory followed by the model when we change the lattice couplings. The dashed purple line shows the critical surface separating distinct basins of attraction. In the lattice simulations, we expect to observe a behavior change when we cross this critical surface.
}
\label{fig:RGMap}
\end{figure}

Looking for a second order phase transition in an infinite dimensional space of couplings is a field-theoretic version  of the "needle in a haystack" problem. The $\epsilon$ expansion makes it not completely hopeless, as it suggests that,  in the sense of the RG and in five dimensions, only one operator is relevant. This supports the existence of a critical surface separating two phases which can be reached by   tuning only  one coupling constant. From the perspective of lattice models, we expect that starting from an arbitrary point in the space of couplings, one should in principle be able to reach this critical surface by scanning a one-dimensional space. We show a sketch of this scenario in Fig.~\ref{fig:RGMap} for three couplings $g_1,g_2,g_3$. We pick $g_1$ and $g_2$ to be irrelevant couplings and $g_3$ to be the relevant direction. We denote the conjectured fixed point by a black dot and draw the attractive and repulsive directions. The critical surface is generated by the two irrelevant directions and is depicted by dashed lines. In black, we show a potential trajectory (obtained varying only one parameter in this case, for the clarity of the drawing) in the coupling space which could potentially be successfully followed to discover the existence of the critical surface and its associated critical point. For completeness, we also draw in gray the would be "continuum limit" which one could take to define a valid continuum theory.

Keeping this picture in mind, we will study two distinct lattice models, whose actions depend on two parameters. In particular, we will study their phase space. After reviewing our notations, we describe the specific models in the next section. A second crucial point will be to be able to measure with relative certainty the (non)-existence of a second order phase transition. As already discussed in previous work, see for instance \cite{Creutz:1979dw,Kawai:1992um}, it can be very challenging to distinguish between a weakly first order phase transition and a true second order transition. Moreover, as we will show in Sec.~\ref{subsec:resfa}, a vanishing "latent heat" \footnote{The discontinuity in the average value of the  action.} is not a sufficient condition to guarantee the existence of a second order phase transition. To overcome this issue, we will use Monte-Carlo techniques borrowed from condensed matter \cite{velazquezExtendedCanonicalMonte2016}, not so dissimilar in spirit to some multicanonical algorithms that were for instance used to study the topological charge in lattice QCD \cite{Bonati:2018blm}. We present these techniques in section~\ref{sec:mircocanonical}.

\subsection{Lattice actions for SU(2) in 5 dimensions}

To construct gauge invariant lattice models, we introduce link variables $U^{\mathfrak{R}}_\mu$, which a priori can belong to any representation $\mathfrak{R}$ of $SU(2)$. As usual, gauge invariant objects can only\footnote{In theories with matter in the fundamental representation like quarks, it is possible to obtain distinct gauge invariant objects by attaching quarks to the endpoints of Wilson lines.} be obtained by tracing over closed loops made out of these links. These  "Wilson loops" $\square_{\mu\nu, I\times J}^\mathcal{R}$ are ordered products of links $U^{\mathfrak{R}}_\mu$ in the plane $\mu,\nu$, making a rectangle of size $I\times J$.  As we will only consider isotropic systems, we will often omit the directional index $\mu$ and $\nu$ and simply refer to Wilson loops of size $I$ by $J$ by $\square_{I\times J}^\mathcal{R}$.

The first action we study is
\begin{align}
\label{Fundamental + Adjoint Action}
S^{f,a}_{1\times 1}=\sum_{\square_{1\times 1}}\dfrac{\beta_{f}}{2}\text{Tr}\left(1 - \square^f_{1\times 1}\right)+\dfrac{\beta_{a}}{3}\text{Tr}\left(1 - \square^a_{1\times 1}\right) \ ,
\end{align}
where the sum is performed over all distinct plaquettes.
This model is an ideal starting point. Indeed, it has already been studied in \cite{Kawai:1992um}, where some phase transitions could not be well resolved, leaving open the possibility of them being second order. As we will show in the next section, this is unfortunately not the case. This led us to consider a second model, defined by the action
 \begin{align}
 \label{2 Loops Action}
S^f_{1\times 1, 2\times 2}=\sum_{\square_{1\times 1}}\dfrac{\beta_{1}}{2}\text{Tr}\left(1 - \square^f_{1\times 1}\right)+\sum_{\square_{2\times 2}}\dfrac{\beta_{2}}{2}\text{Tr}\left(1 - \square^f_{2\times 2}\right) \ .
\end{align}
A compelling reason to believe this second model explores a different region in coupling space is the absence of an analytical relation between the 1-sided and 2-sided loops. This is in contrast to what happens (in the case of $SU(2)$ and as recalled in appendix~\ref{app:SU2RepTheory}) in the fundamental plus adjoint extension. There, the different traces are related by
\[
\operatorname{Tr}(\square^a_{I\times J}) = \operatorname{Tr}(\square^f_{I\times J})^2 - 1 \ .
\label{AfromF}
\]

It is also worth noting that all of these models have the same naive continuum limit; all different operators tend to  $\alpha^\mathfrak{R}_{I\times J}\Tr{G_{\mu\nu}G^{\mu\nu}}$ with $G_{\mu\nu}$ a Lie algebra valued field-strength tensor and $\alpha^\mathfrak{R}_{I\times J}$ is a constant which depends on the loop size and representation. In particular, one can explicitly check that
 \begin{align}
 S^{f,a}_{1\times 1}& \to \int \dd x^5 \frac{1}{2g^2}\Tr{G_{\mu\nu}G^{\mu\nu}}  \ ,\qquad
 \qquad  \dfrac{4}{g^2}=\beta_f + \dfrac{8}{3}\beta_a, \\
S^f_{1\times 1, 2\times 2}&\to\int \dd x^5 \frac{1}{2g^2}\Tr{G_{\mu\nu}G^{\mu\nu}}  \ ,\qquad
 \qquad  \dfrac{4}{g^2}=\beta_1 + 16\beta_2\ .
\end{align}
Before moving on, it will prove convenient to define the following auxiliary quantities, $W^\mathfrak{R}_{I\times J}$,
\[\label{eq:WilsonLoop}
W^\mathfrak{R}_{I\times J} = 1 - \dfrac{\sum_{\square_{I\times J}}\text{Tr}\left(\square_{I\times J}^\mathfrak{R}\right)}{N_\text{loops}d(\mathfrak{R})},
\]
where $N_\text{loops}$ is the number of loops in the lattice\footnote{There are $\genfrac(){0pt}{2}{D}{2} L^D$ loops in a hypercubic lattice of dimension $D$ and side $L$.}, and $d(\mathfrak{R})$ is the dimension of the representation, such that
our actions can be rewritten as

 \begin{align}
 S^{f,a}_{1\times 1}&=N_\text{loops} \left(\beta_{f} W^f_{1\times 1} + \beta_{a} W^a_{1\times 1}\right ) \\
S^f_{1\times 1, 2\times 2}&=N_\text{loops}\left (\beta_{1}W^f_{1\times 1} +\beta_{2}W^f_{2\times 2} \right ) \ .
\end{align}

\subsection{Phase transitions from microcanonical measurements}\label{sec:mircocanonical}

We aim to search for a second order phase transitions in the two models presented in the previous section.
 A conventional way to proceed would be to analyze the free energy of the system as a function of our coupling constants.
 As already mentioned previously, this has the drawback of making it very hard to distinguish between weakly first order and second order phase transitions. This difficulty arises from the fact that a first order phase transition is characterized by a discontinuity in the derivative of the free energy. The smaller the discontinuous jump is, the weaker the transition is. Resolving small discontinuities require increasingly more precision in the region of parameter space around the phase transition. Unfortunately this is precisely the region where Monte-Carlo simulations become increasingly harder to perform (because of the existence of different disconnected phases in the first order case and of critical slowing down in the second order case, see for instance \cite{Creutz:1979dw}).
 This method is inspired by multicanonical methods, as in both cases we control the sampled distribution to improve the estimates of the relevant quantities (see for instance \cite{PhysRevE.74.046702,Berg:1992qua,PhysRevLett.71.211,PhysRevLett.63.1195}).

We will show now that this problem can be evaded by considering modified ensemble weights.
The intuition behind this idea is the following. A first order transition corresponds to a point in coupling space with distinct global minima for the free energy (separated by local maxima). Just after the phase transition, one of the minima becomes the "true" one, but a system with a fluctuating action, a "canonical ensemble" in statistical mechanics terms, will have trouble thermalizing.
This leads to a phenomenon known as hysteresis cycles, which is also aggravated by the lattice size. By changing the weights, we can control the number of minima. With these modified weights, we are not  able to directly calculate averages, as is usually done. However, we are still able to estimate the microcanonical entropy, which contains enough information to identify the order of the phase transitions.

More concretely, let us rewrite our standard ("canonical") partition function as a sum of contributions coming from fixed actions ("microcanonical")

\begin{align}
Z=\sum_\Sigma e^{-S[\Sigma]} \equiv \sum_S e^{-S} \rho(S) \ , \label{eq:partitionFuncCanonicalMicro}
\end{align}
with $\Sigma$ denoting a field realization and $\rho(S)$ is a "density of states"; it counts the number of states with action $S$. In particular, noting that the Boltzmann entropy $S_B$ is simply given by the logarithm of $\rho$
\[
S_B(S)=\log(\rho(S)) \ ,
\]
we can rewrite Eq.~\ref{eq:partitionFuncCanonicalMicro} as
\[
Z=\sum_S e^{-S + S_B(S)} \label{eq:partitionFuncCanonicalMicroBis} \ .
\]

We can clearly see from  Eqs.~\ref{eq:partitionFuncCanonicalMicro} and \ref{eq:partitionFuncCanonicalMicroBis} that all the analytic structure of $Z$  and hence the thermodynamics of the system is encoded in $\rho(S)$ or equivalently in the entropy $S_B(S)$. This fact is crucial, as we will now explain how $S_B$ can reliably be measured even in the vicinity of a phase transition, following the method presented in \cite{velazquezExtendedCanonicalMonte2016}.

Let us start by recalling how $S_B(S)$ can be measured from standard simulations. The main step is to compute the probability density \[p(S)\equiv e^{-S} \rho(S) \ .  \label{eq:probDens}\] Up to a normalization constant $C$, this is done  by generating a histogram of the Monte Carlo chain  binned by action, $H(S)$. One then can recover the entropy $S_B$
\[S_B(S)=\log\left(H(S)\right)+S + \log(C) \ .\]
In practice, it is often more convenient to directly consider derivatives of the above quantities with respect to the action, so that the normalization factor drops.

Now consider the case  at hand where the action is of the type $S=N_{loops} x$, with $x$ some arbitrary variable. Eq.~\ref{eq:probDens} then reads
\[p\left(x\right)= e^{-N_\text{Loops} x} \rho\left(x\right) \ \ .\]
Moreover, were we to change the ensemble weights (the weight given to each configuration) from the action $\exp\left( -S\left[x\right]\right) \to \exp\left( -S'\left[x\right]\right)$ we would only affect the probability $p\left(x\right)\to p'\left(x\right)$, but not the density of states $\rho\left(x\right)$. If we only want to measure the density of states, we can now change the ensemble weights, to optimize our analysis. It is useful to parametrize this new function as
\[\label{eq:new_ensemble_omega}
S'\left[x\right] = \omega\left(x\right)x N_\text{loops}.
\]
In particular, we will choose
\[\label{eq:new_ensemble}
\omega\left(x\right) = (\omega_2 x+\omega_1)
\]
with $2\omega_2>\max\left(\frac{d^2 S_B}{d x^2}\right)/N_\text{Loops}$.

This choice greatly improves  performances over canonical ensemble Monte Carlo simulations close to the phase transition.
This can be understood as follows. The main contribution to the integral in  Eq~\ref{eq:partitionFuncCanonicalMicroBis} comes from the global maximum of the integrand, obtained by minimizing the exponent (which is the statement that systems at equilibrium are at the minimum of the free energy)
\[\label{eq:maximum}
\dfrac{d p\left(x\right)}{dx} = 0 \Longrightarrow   \dfrac{d S_B}{dx} = \dfrac{d S^\prime\left[x\right]}{dx}.
\]
A minimum does of course correspond to solutions with negative second derivatives. Using the modified ensemble  \eqref{eq:new_ensemble}, the Eq.~\eqref{eq:maximum} reduces to
\[
\label{eq:simpmaximum}
\dfrac{d S_B}{dx} =\left( 2\omega_2x + \omega_1 \right) N_\text{loops}\ .
\]

\begin{figure}
\centering
\includegraphics[width=0.9\textwidth]{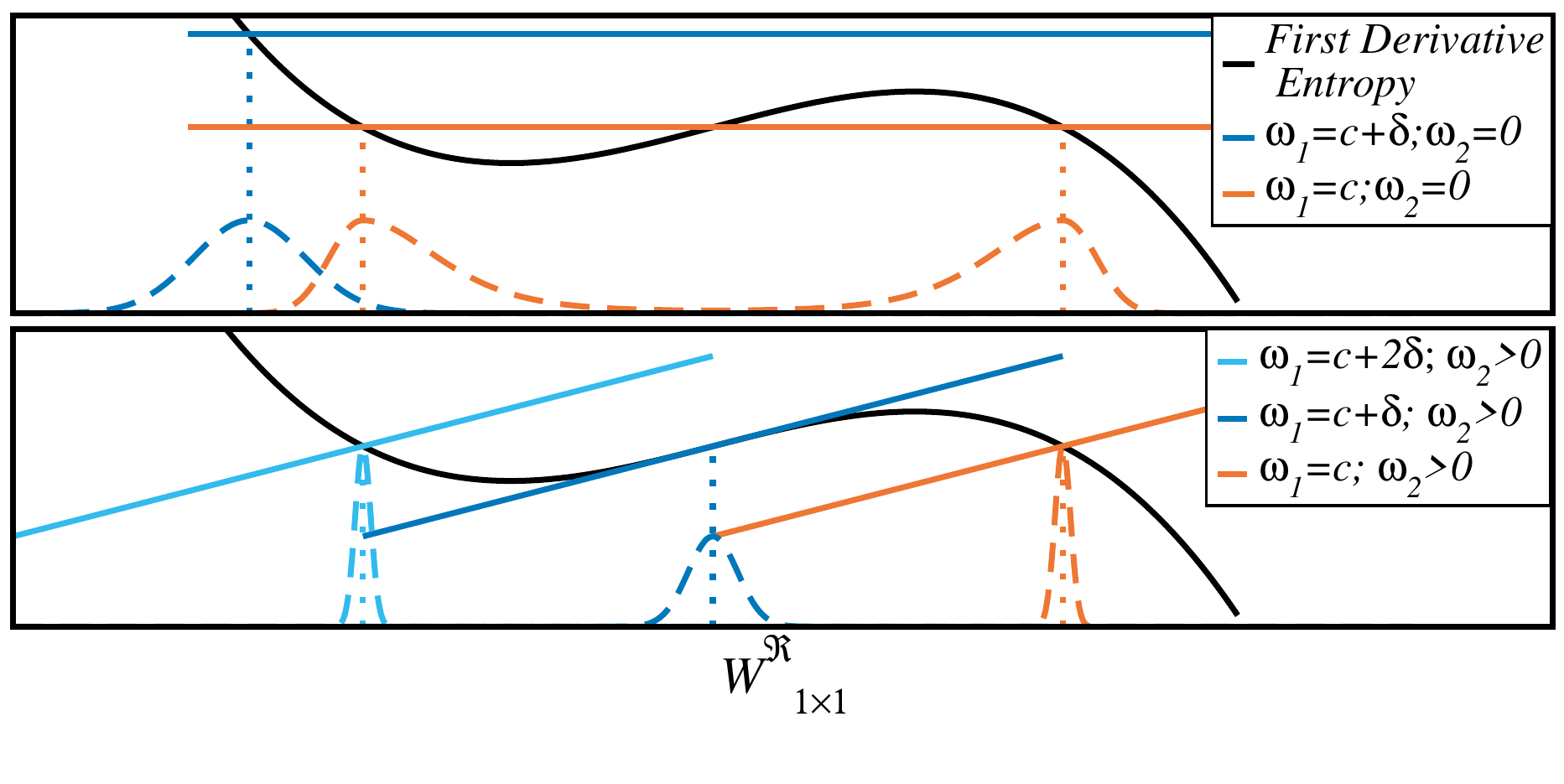}
\caption[A]{Schematic representations of Eq.~\ref{eq:maximum}, for distinct ensembles. The black lines represent the first derivative of the entropy near a first order phase transition. The straight blue ones show the contributions coming from the different ensembles. On the bottom of each plot, we sketch the histograms expected when performing Monte Carlo with these parameters. In the \textbf{top}, we present the canonical ensemble. Depending on the value of the $\omega_1$, we can have more than one solution, as seen in the lighter curve. A slight change in the parameter will transform one of the global maximum into a local maximum (of the probability). %Changing it even more will destroy this local maximum, as seen in the darker blue line and we will have only a maximum (global or local).
It is this local maximum that leads to hysteresis. On the \textbf{bottom}, we show the results for the ensemble proposed in Eq.~\ref{eq:new_ensemble}. For a large enough value of $\omega_2$, there will be only a solution, for all values of $\omega_1$, which removes hysteresis cycles.
}
\label{fig:EntropyScheme}
\end{figure}

We sketch the  behavior of the first derivative of the entropy close to a first order phase transition, in Fig.~\ref{fig:EntropyScheme}. In black, we depict the generic behavior of this quantity around a first order phase transition. The straight lines are the right hand-side of Eq.~\ref{eq:simpmaximum} when $\omega_2 =0$ (canonical ensemble); the intersection between these curves thus extremize the probability distribution. In this case, the existence of a first order phase transition manifests itself as the existence of two separate minima (in the intersection point, the slope of the first derivative of the entropy is negative) of the entropy. Physically, this simply corresponds two the fact that both phases coexist at the phase transition. The existence of these separate minima is what makes standard Monte-Carlo simulations hardly tractable close to phase transitions, as the Monte-Carlo chain will tend to get stuck in one of the two phases. This leads to typical "hysteresis cycles", where the obtained results depend on whether the phase transition was approached from above or from below. In terms of entropy, this can be easily identified as a region with a positive second derivative.

We can now understand the impact of $\omega_2\neq 0$. In particular, when $2\omega_2>\max\left(\frac{d^2 S_B}{d x^2}\right)/N_\text{Loops}$, Eq.~\ref{eq:simpmaximum} admits only a single solution; this is what is depicted in the bottom panel of Fig.~\ref{fig:EntropyScheme}. As a result, Monte-Carlo simulations with these modified ensemble weights do not suffer from the problems associated with the first order phase transition anymore (hysteresis cycles and unsamplable values of the action).

The models used in this study have more than one parameter - there is a different coupling associated with each contribution to the action. In this case, the microcanonical entropy depends on both contributions
\[
S_B=S_B(W_{1\times 1}^f,W_{I\times J}^\mathfrak{R}) \ ,
\]
where in our case we have $W_{I\times J}^\mathfrak{R}=W_{1\times 1}^a$ or $W_{I\times J}^\mathfrak{R}=W_{2\times 2}^f$ depending on the model. %In this case,
For a given amount of statistics, this greatly degrades the quality of the results because now the histogram used to measure the entropy is two-dimensional. To mitigate this problem, we define a reduced entropy %$S_B(W_{1\times 1}^f)$
\[
e^{S_B(W_{1\times 1}^f)} = \int \dd W_{I\times J}^\mathfrak{R} \exp{\left(S_B(W_{1\times 1}^f,W_{I\times J}^\mathfrak{R}) - \beta\, W_{I\times J}^\mathfrak{R} N_\text{loops}\right)},
\]
where $\beta = \beta_a$ if $W_{I\times J}^\mathfrak{R}=W_{1\times 1}^a$ or $\beta =  \beta_2$ if $W_{I\times J}^\mathfrak{R}=W_{2\times 2}^f$.
From now on, we will refer to $
S_B(W_{1\times 1}^f)$ as the entropy.
We shall use the deformed ensemble method explained above only in the variable $x=W_{1\times 1}^f$.
Improving the signal comes at the cost of losing some information in the projection. Since we are only looking for thermodynamics discontinuities characteristic of phase transition, they will manifest themselves also in the projected quantity. The only potential drawback is that they may be harder to distinguish before the infinite volume limit ($L\to \infty$)  is reached. While this is a price we need to pay, this did not seem to happen in our system, as we will show in the next section.

Before moving on to the results, let us mention that the interested reader can read more in App.~\ref{App:MonteCarloAlgorithms} about how the standard Monte-Carlo algorithms need to be modified to work with our modified action.

\subsection{Predicted scaling by the $\epsilon$ expansion}
\label{sec:prediction}
The $\epsilon$-expansion  not only predicts the existence of RG fixed point but also allows for the estimation of its critical exponents. Based on this result, we can infer the expected behavior of the model near criticality. In this subsection we will use general scaling arguments to predict the behavior of the specific heat near the critical point.
Let us call $u=\beta - \beta_c$ the coupling we vary across the phase transition (the others are kept fixed). Let us call $W$ the term that multiplies $u$ in the action. Then, we can write
\begin{align}
e^{-N f(u)} = \int dW e^{N ( s_B(W)-u W )}
\label{thermo}
\end{align}
where $f$ and $s$ are the free energy and entropy per site.
From standard scaling arguments (see for instance \cite{Cardy:1996xt}), we find that the singular part of the free energy scales as
\begin{align}\label{scaling free energy}
    f(u) \sim   |u|^{d \nu}\, ,
\end{align}
where $d=5$ is the spacetime dimension and  the critical exponent  $\nu \approx 0.62 $ can be estimated in the epsilon expansion
\cite{Morris:2004mg, DeCesare:2021pfb}.
From \eqref{thermo} in the thermodynamic limit, we find
\begin{align}
    \left\langle W\right\rangle &= \dfrac{df}{du} = W_c+ O(u) +   O\left( |u|^{ d\nu -1}\right)
    \label{average value}
\end{align}
At this point, it is important to distinguish two cases: either $d\nu<2$ or $d\nu>2$. Let us first consider $d\nu<2$, such that $ |u| \sim |W-W_c|^{\frac{1}{d\nu-1}}$. Using \eqref{scaling free energy} and \eqref{average value}, we can obtain the scaling of the entropy near criticality
\begin{align}
      s_B &= u \left\langle W\right\rangle -f \\
      & \sim f_c+O(u^2) +  O\left(|u|^{ d\nu }\right)\\
      & \sim s_{B_c}+ O\left(|W-W_c|^\frac{2}{ d\nu -1}\right)+   O\left(|W-W_c|^\frac{d\nu}{ d\nu -1}\right)
\end{align}
In this case,  the second term dominates when $W \to W_c$, hence the second derivative of the entropy vanishes, \footnote{Notice that the unitary bounds imply $d\nu>1$.}
\begin{align}
   \frac{d^2 s_B}{dW^2}
   \sim  |W-W_c|^\frac{2-d\nu}{ d\nu -1}
\end{align}
and the specific heat diverges. The 3D Ising model is an example of this class.

Consider now the case $d\nu>2$, such that $ |u| \sim |W-W_c|$ and
\begin{align}
      s_B &= u  \frac{df}{du} -f - \text{const} \sim \text{regular} +  O\left(|u|^{ d\nu }\right) \sim \text{regular} +   O\left(|W-W_c|^{d\nu}\right)
\end{align}
In this case, the second derivative of the entropy is dominated by the regular term, %\footnote{Notice that the unitary bounds imply $d\nu>1$.}
\begin{align}
   \frac{d^2 s_B}{dW^2}
   \sim \text{regular} + |W-W_c|^{d\nu-2}
\end{align}
and the specific heat generically does not diverge. The O(2) model in 3D is an example of this case. The behavior of the specific heat is famously known as the lambda point of the helium superfluid transition.

The $\epsilon$-expansion estimates of the recent paper \cite{DeCesare:2021pfb}
suggest that 5D YM falls in the second case.

\section{Results}

We  start in Sec.~\ref{subsec:resfa} by presenting results obtained for our first model \eqref{Fundamental + Adjoint Action} made of fundamental and adjoint plaquettes. The main motivation behind this choice is the existence of a previous study \cite{Kawai:1992um} where second order criticality could not be excluded in some region of the parameter space. We first chart this parameter space for small lattices and locate the same region of interest as in \cite{Kawai:1992um}. After characterizing the nature of the different phases, we  perform a finite size scaling study and show that a second order phase transition is disfavored by our new datasets and improved algorithms.

\subsection{Fundamental + Adjoint Action}
\label{subsec:resfa}

\begin{figure}
\centering
\includegraphics[width=0.45 \textwidth]{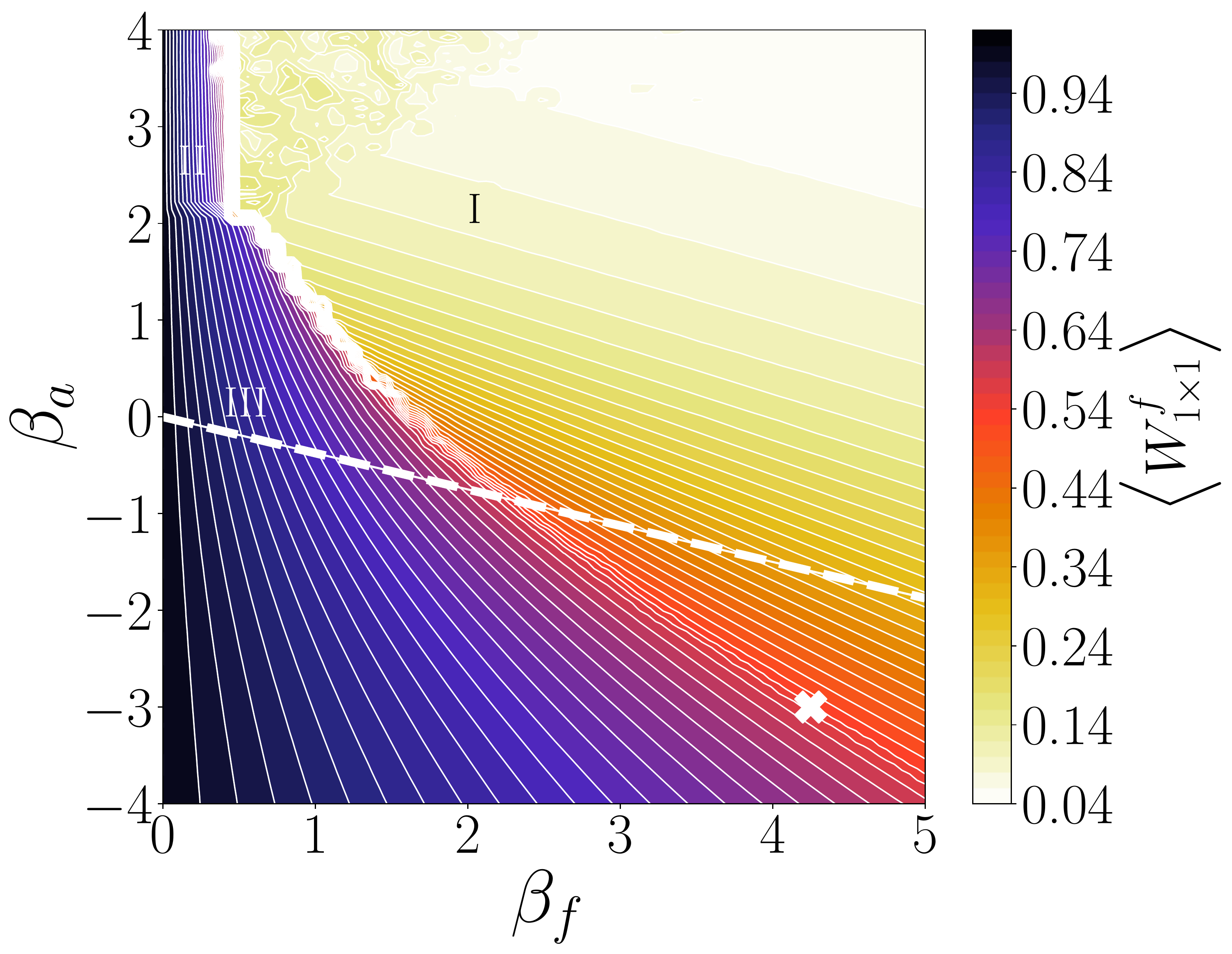}\includegraphics[width=0.45 \textwidth]{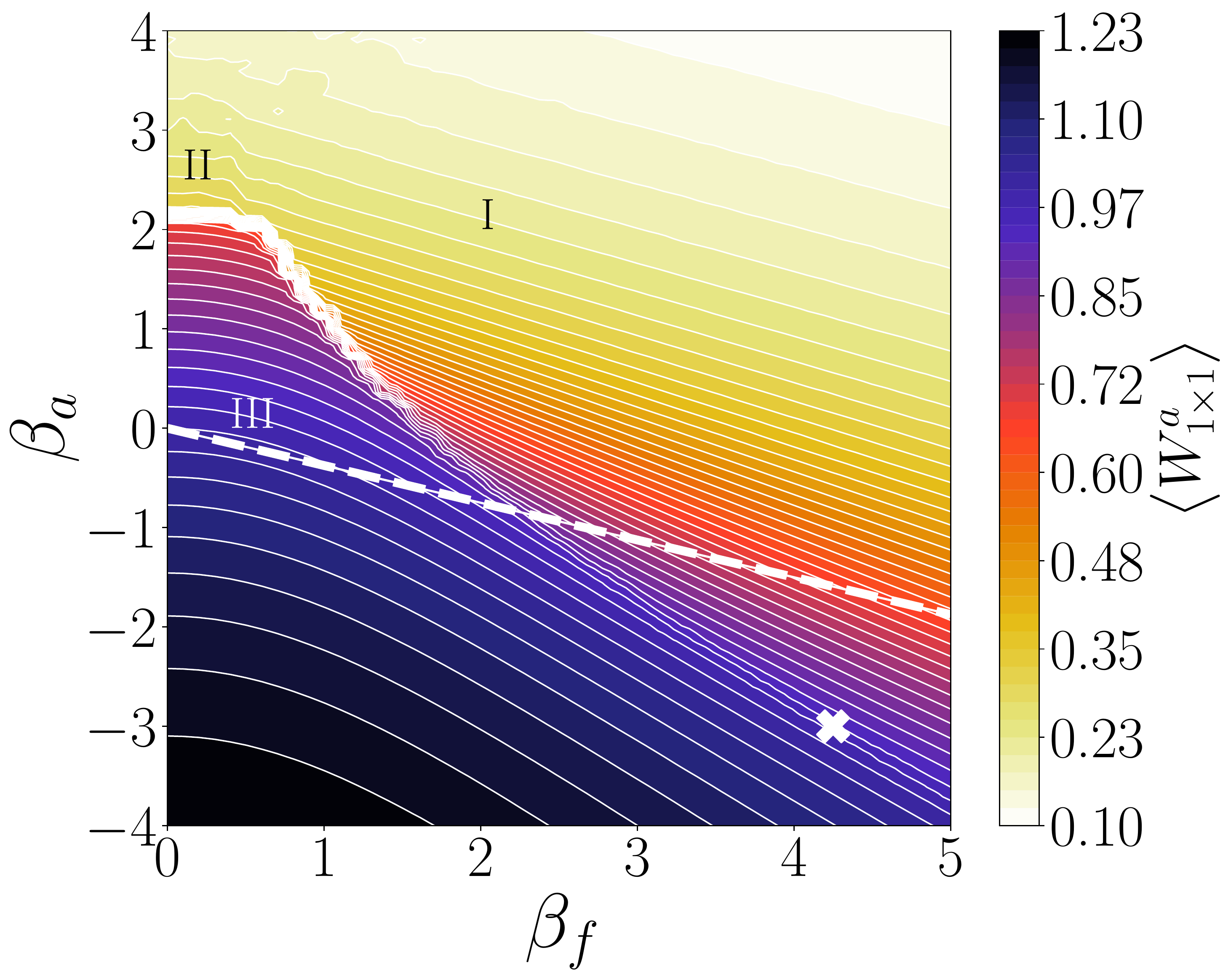}
\caption[Phase Space Scan]{Survey of the parameter space, for $L=4$. The image on the \textbf{left} represents $\left\langle W_{1\times 1}^f \right\rangle _{\beta_{f},\beta_{a}}$ and the one on the \textbf{right} $\left\langle  W_{1\times 1}^a\right\rangle _{\beta_{f},\beta_{a}}$. The white cross marks a point discussed in \cite{Kawai:1992um}. We identified three different "phases", labeled by the roman-numerals. The thin white lines are level curves, therefore the closer they are, the sharper the change in value is. We identify discontinuities in the sampled values, when several white lines come together. The dashed white line distinguishes regions with different signs of the coupling in the naive continuum limit; below  $\nicefrac{1}{g^2} < 0$ and above $\nicefrac{1}{g^2} > 0$ . We used rectangular grid with a spacing of $0.1$.}
\label{fig:PhaseSpaceScan}
\end{figure}

To search for potential second order phase transitions, we need to get some understanding of the phase diagram. With this in mind, we choose $L=4$ lattices and sample $W^f_{1\times 1}$ and $W^a_{1\times 1}$ in a selected region of the parameter space $(\beta_a, \beta_f)$ around zero. Actually, only the $\beta_f>0$ region needs to be explored as the action for negative $\beta_f$ is not independent; the two regions are related by the following reflection formulae
\[
\left\langle W^f_{1\times 1} \right\rangle _{\beta_{f},\beta_{a}}+\left\langle W^f_{1\times 1}\right\rangle _{-\beta_{f},\beta_{a}}=2,\qquad \qquad \left\langle W^a_{1\times 1}\right\rangle _{\beta_{f},\beta_{a}}=\left\langle W^a_{1\times 1}\right\rangle _{-\beta_{f},\beta_{a}} \ . \label{eq:reflection_formula}
\]
This follows from the symmetry of the action (up to an irrelevant constant)
\begin{equation}
\beta_f \to -\beta_f\,,\qquad\qquad
U_\mu^f (x)   \to
(-1)^{\sum_{\alpha=1}^{\mu-1} x^\alpha}\
U_\mu^f (x)  \,,
\label{eq:sym}
\end{equation}
where we measure distances in lattice units so that the coordinates $x^\alpha \in \mathbb{Z}$.
Notice that this transformation changes the sign of all fundamental plaquettes $ \operatorname{Tr}(\square_{1\times 1}^f)$, leaving the adjoint plaquettes invariant as $\operatorname{Tr}(\square^a_{1\times 1}) = \operatorname{Tr}(\square^f_{1\times 1})^2 - 1$, see also \cite{Li:2004bw} for a similar method.

We present the phase diagram we obtain in Fig.~\ref{fig:PhaseSpaceScan}, where $W^f_{1\times 1}$ is represented on the left hand-side and $W^a_{1\times 1}$ in the right hand-side. To obtain these results, we used standard "canonical simulations". The discontinuities in these two quantities, seen as an accumulation of level lines (white lines), allow us to identify three different phases, denoted by I, II, III in Fig.~\ref{fig:PhaseSpaceScan}. For $\beta_f = 0$, we expect to see a strong first order phase transition for the $SO(3)$ model \cite{Baig:1984rg} with a critical temperature of around $\beta_a\approx 2$. For $\beta_a = 0$, we expect the two usual confined/deconfined  phases, see for instance \cite{Kawai:1992um}.
Phases I and II are separated by another first order phase transition. At large $\beta_a$, the adjoint action restricts the ordered product of the links to be 1 or -1, which in turn requires the links themselves to be 1 or -1 up to gauge transformations. By restricting the system to the ground state of the adjoint action, the only fluctuating term corresponds to the Wilson action (first term on the right hand-side of \eqref{Fundamental + Adjoint Action}) of the $\mathbb{Z}_2$ gauge group (the links can either be 1 or -1), which is known to have a first order phase in 5 dimensions, at $\beta_f=0.35$ \cite{Kawai:1992um}.

As we are interested in phase transitions, we focused on the boundaries between these regions. In particular, we are looking for points for which the first order phase transition is becoming weaker (smoother color gradients). The only region meeting these criteria is identified by a white cross, which is precisely the region already investigated in \cite{Kawai:1992um}. Having verified that all the other boundaries are indeed first order phase transitions, we focus our efforts on this region.
\begin{figure}
\centering
%\begin{mdframed}[backgroundcolor=blue!50,linecolor=blue!50]
\centering
\includegraphics[width=0.5 \textwidth]{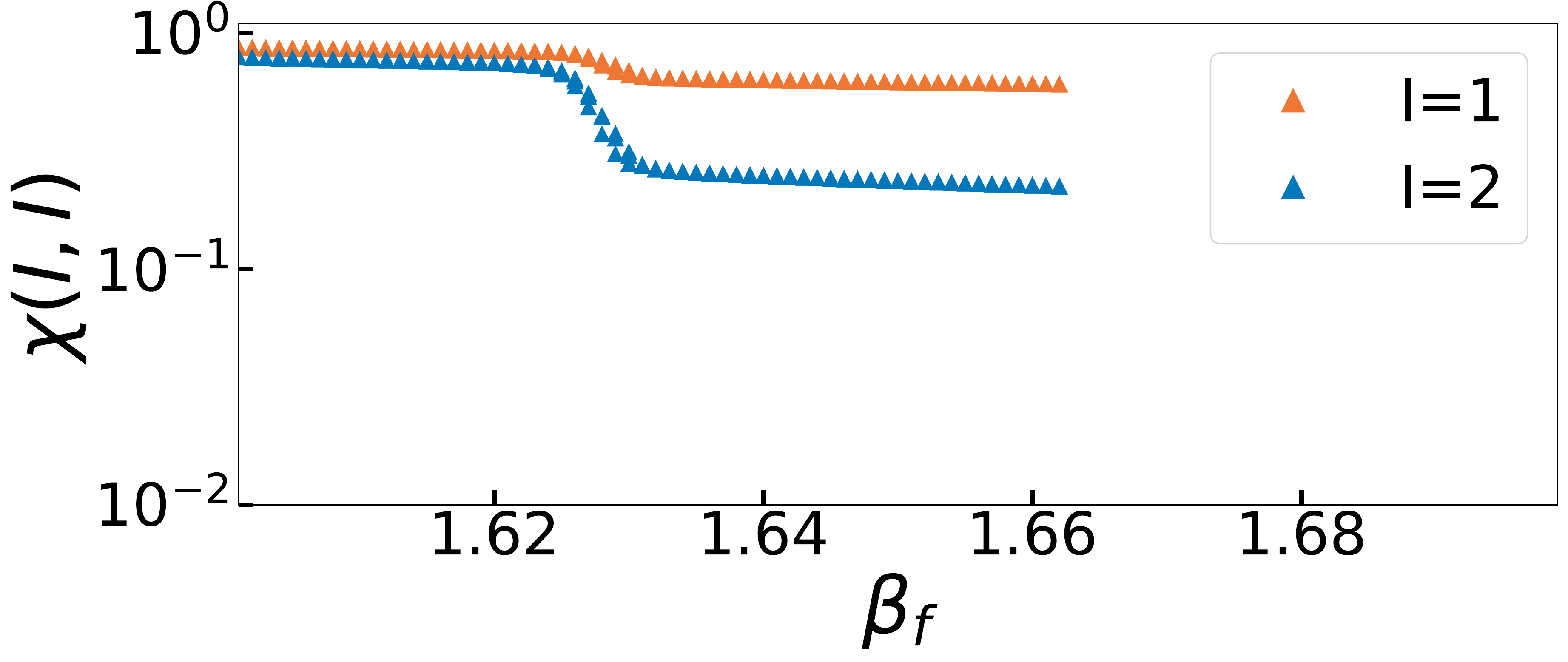}\includegraphics[width=0.5 \textwidth]{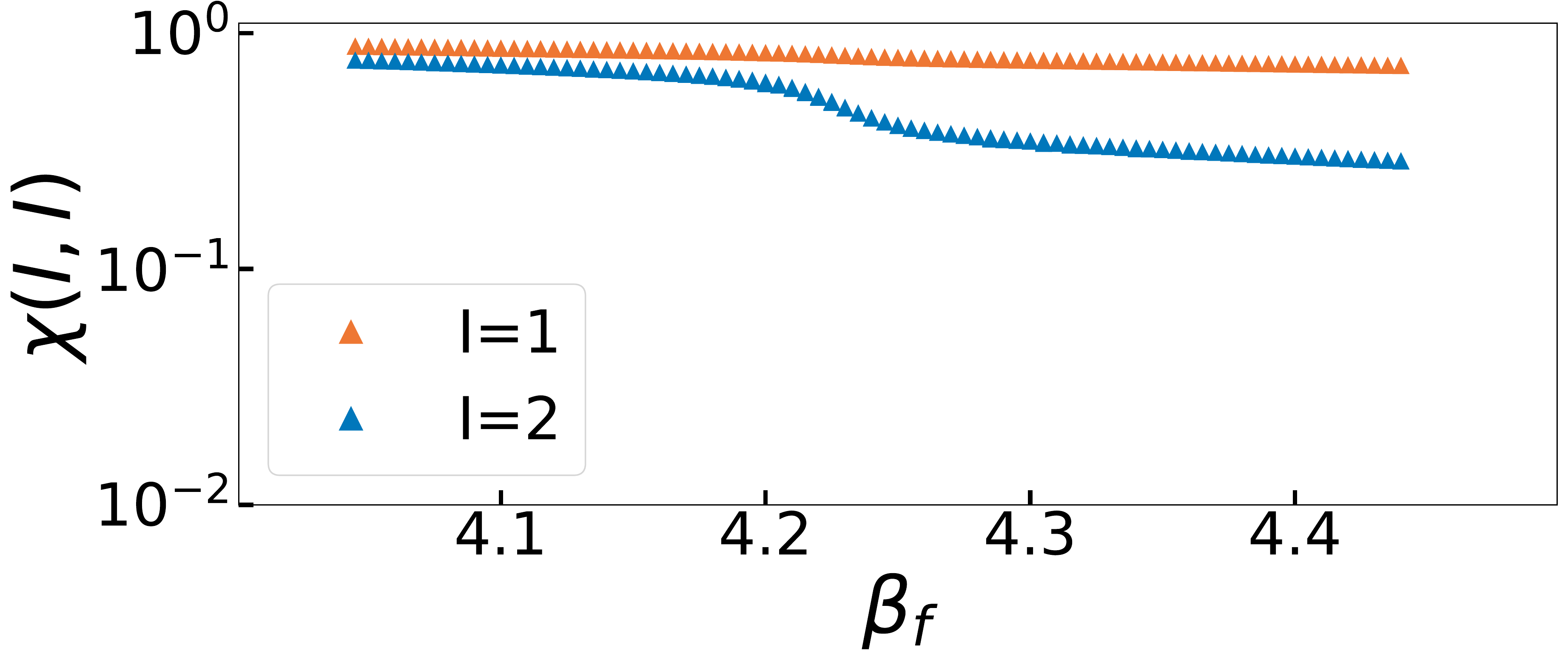}
\includegraphics[width=0.5 \textwidth]{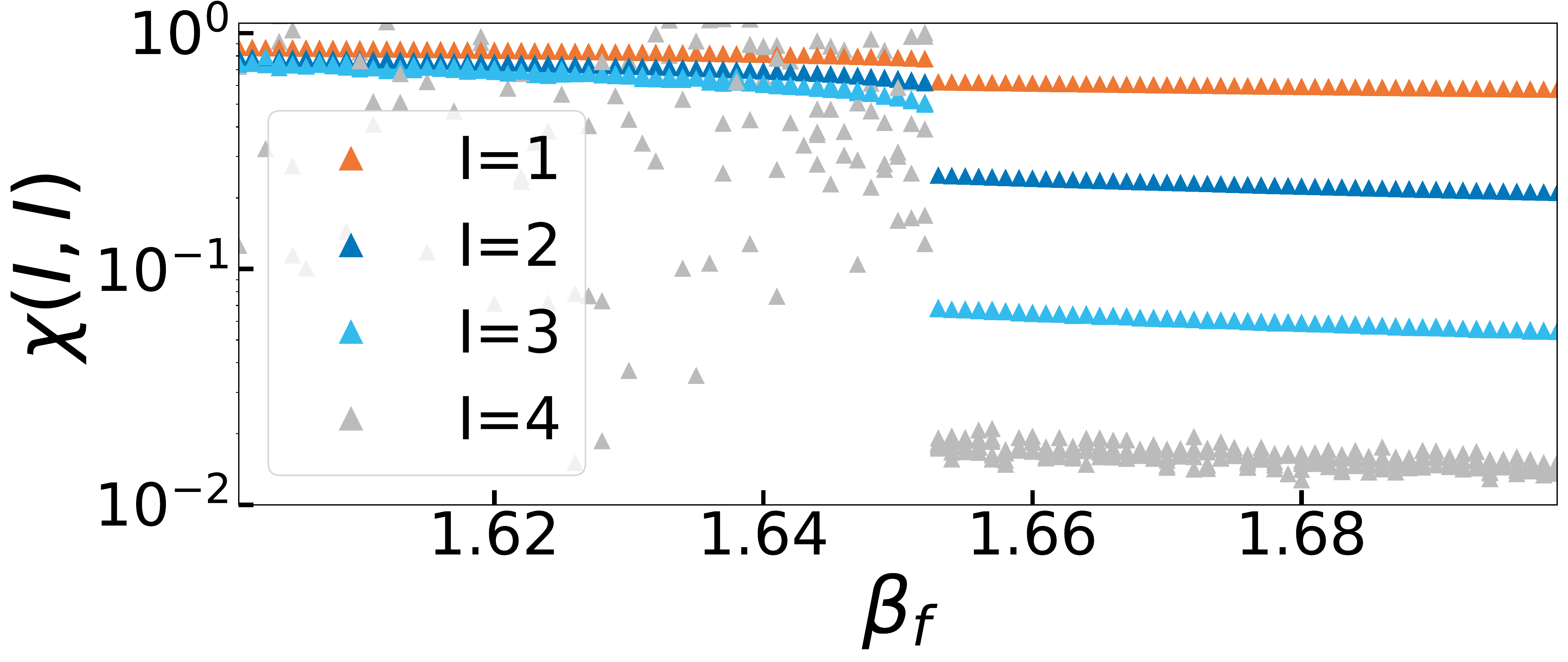}\includegraphics[width=0.5 \textwidth]{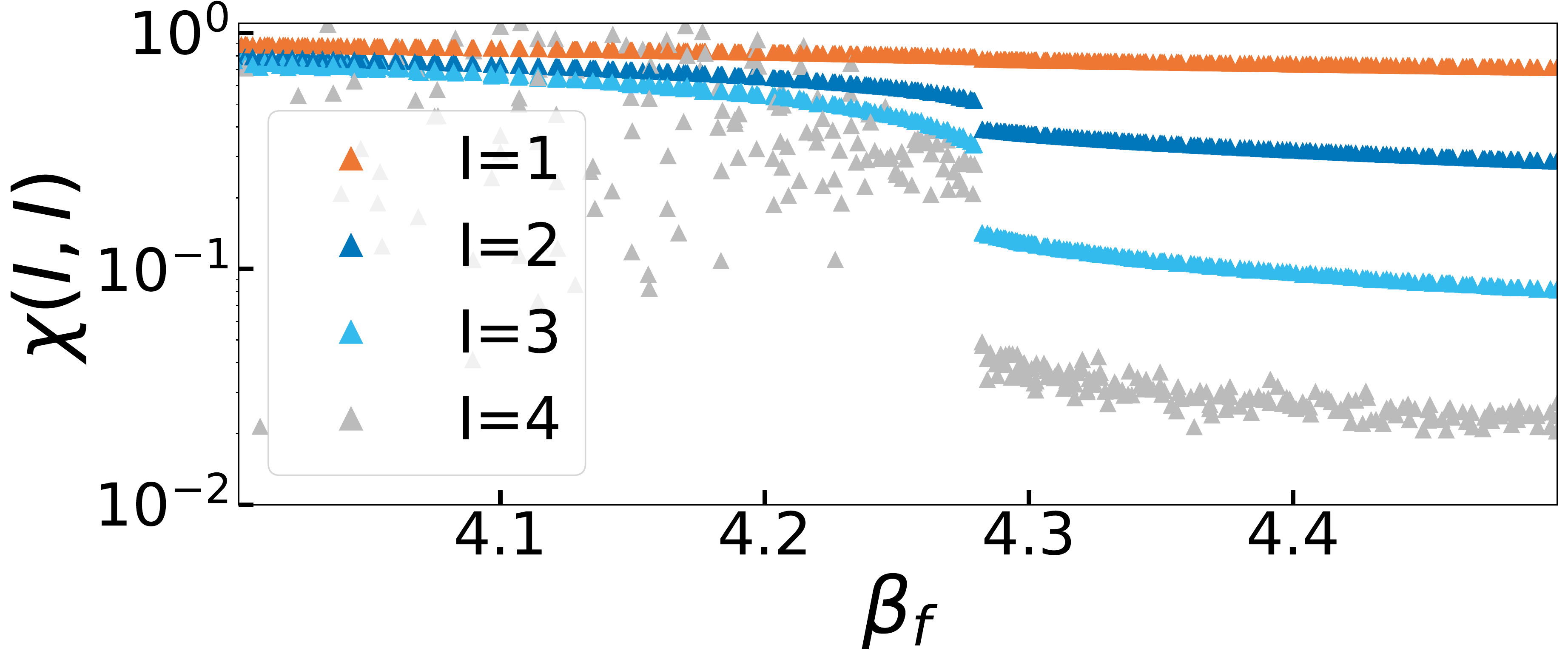}
\caption{Creutz's ratios for $\beta_a=0$ (\textbf{left}) and $\beta_a=-3$ (\textbf{right}), $L=4$ (\textbf{top}) and $L=8$ (\textbf{bottom}). We present the results in a logarithm scale. In the deconfined phase, the Creutz ratio goes to zero in the limit of large Wilson loops. %We use a logarithmic scale to show the almost exponential decay of the Creutz ratios, for the first loop sizes.
For the plots on the left hand-side, the error vary between $10^{-7}$ and $10^{-3}$. For the plots on the right hand-side we show the results for 5 independent simulations (with distinct values of $\beta_f$), whose spread gives us a good estimate for the error. The errors are significantly smaller than the markers, except for $I=4$ especially in the confined phase. This arises because the Creutz ratio is a ratio between very small numbers. The Wilson Loops in the confined phase are around $10^{-7}$, compared with $10^{-4}$ in the deconfined phase.  }
\label{fig:CreutzRatios}
%\end{mdframed}

%\centering
%\includegraphics[width=0.5 \textwidth]{Chi_Beta0L4.pdf}\includegraphics[width=0.5 \textwidth]{Chi_Beta-3L4.pdf}
%\includegraphics[width=0.5 \textwidth]{Chi_Beta0L8.pdf}\includegraphics[width=0.5 \textwidth]{Chi_Beta-3L8.pdf}
%\caption{Creutz's ratios for $\beta_a=0$ (\textbf{left}) and $\beta_a=-3$ (\textbf{right}), $L=4$ (\textbf{top}) and $L=8$ (\textbf{bottom}). We present the results in a logarithm scale. In the deconfined phase, the Creutz ratio goes to zero in the limit of large Wilson loops. %We use a logarithmic scale to show the almost exponential decay of the Creutz ratios, for the first loop sizes.For the plots on the left hand-side, the error vary between $10^{-7}$ and $10^{-3}$. For the plots on the right hand-side we show the results for 5 independent simulations (with distinct values of $\beta_f$), whose spread gives us a good estimate for the error. The errors are significantly smaller than the markers, except for $I=4$ especially in the confined phase. This arises because the Creutz ratio is a ratio between very small numbers. The Wilson Loops in the confined phase are around $e^{-16}$, compared with $e^{-8}$ in the deconfined phase.  }
%\label{fig:CreutzRatios}
\end{figure}

We start by characterizing phases I and III, which are on the two sides of the candidate second order phase transition. Building on previous results \cite{Creutz:1979dw,Kawai:1992um}, we expect a confinement/deconfinement phase transition. As is well-known, see \cite{Creutz:1984mg} for a review, in a pure gauge theory, confinement versus deconfinement can be characterized by the asymptotic behavior of the Wilson loops. In a confined phase, they are expected to follow an area law $\left\langle \operatorname{Tr} \left(\square^f_{I\times J}\right)\right\rangle\sim e^{-KIJ}$ with some constant $K$, interpreted as the string tension, while in the deconfined phase they should follow a perimeter law and decay as  $\left\langle \operatorname{Tr} \left(\square^f_{I\times J}\right)\right\rangle \sim e^{-k(I+J)}$ with some other constant $k$. This information can be neatly rearranged into the so-called "Creutz ratios" \cite{Creutz:1980wj}

\[
\chi(I,I) = -\ln \left(\frac{\left\langle \operatorname{Tr} \left(\square^f_{I\times I}\right)\right\rangle \left\langle \operatorname{Tr} \left(\square^f_{(I-1)\times (I-1)}\right)\right\rangle}{\left\langle \operatorname{Tr} \left(\square^f_{I\times (I-1)}\right)\right\rangle \left\langle \operatorname{Tr} \left(\square^f_{(I-1)\times I}\right)\right\rangle}\right)\sim_{I\to\infty}
\begin{cases}
K & \text{confined phase}\\
0 & \text{deconfined phase}\\
\end{cases}
.\]

To get some understanding of the system's dependence on the lattice size, we start by performing measurements at $\beta_a=0$. The results are shown in the left hand-side of Fig.~\ref{fig:CreutzRatios}. The upper panel shows some Creutz ratios on lattices of size  $L=4$ while the lower panels present results for lattices of size $L=8$. From these plots, we can distiguish two phases; a confined phase with $K$ close to one and a deconfined phase,  with a Creutz ratio approaching zero as a function of the loop size $I$. We see some dependence on the system size but a rather moderate one. The curves overlap, with only a shift in the critical $\beta_f$ value of a few percent.

The right hand-side of Fig.~\ref{fig:CreutzRatios} shows the Creutz ratios obtained for the system with $\beta_a=-3$. As expected from the scan presented in Fig.~\ref{fig:PhaseSpaceScan}, in the case $L=4$ (upper plot) we do not see any phase transition but only what seems to be a smooth crossover between phase III and phase I. As can be deduced from the larger $L=8$ lattices (bottom plot), this is only an artifact of too small volumes; what seems to be a sharp phase transition is observed on the larger system. Also in this case, the transition is between a confined and a deconfined phase.

Having some idea about the nature of the phases under consideration, we now want to move on and explore the phase diagram, looking for a potential second order transition. To this aim, we could  continue using the Creutz ratios as an order parameter. As we will now explain, it is more judicious to use more sensitive tools to better study the nature of the phase transition. Indeed, as we will show, as $\beta_a$ becomes more and more negative, the first order phase transition becomes weaker but does not turn into a second order phase transition. Such a signal is very hard to detect using conventional order parameters.

\begin{figure}
\centering
\includegraphics[width=1 \textwidth]{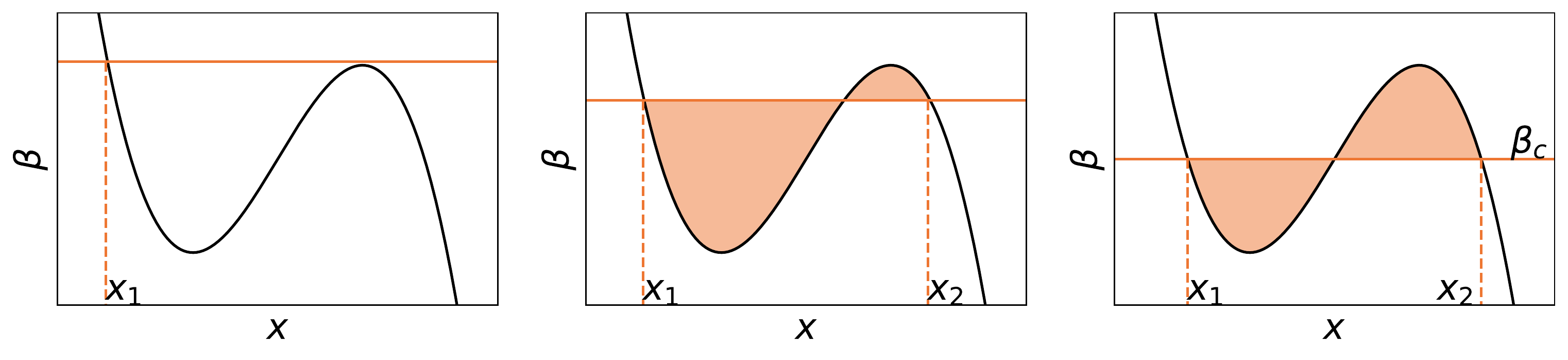}
\caption{Schematic representation of possible choices for $\beta_c$, $x_1$ and $x_2$ satisfying the conditions \eqref{eq:Integral Condition} and \eqref{eq:Maximum Condition}. The black line is a sketch of the first derivative of the entropy near a first order phase transition and the horizontal orange line is $\beta$. The shaded region is the integral in  \eqref{eq:Integral Condition}. Outside the critical region (left), there is only one solution for \eqref{eq:Maximum Condition}. When we approach the phase transition (middle, right) there are 3 solutions for \eqref{eq:Maximum Condition}. We illustrate in the middle panel a choice of $x_1$ and $x_2$ for which for which condition \eqref{eq:Integral Condition} is not satisfied. Actually,  only the  choice of $\beta$, $x_1$ and $x_2$ shown on the right panel satisfies both conditions. This uniquely defines $\beta_c$. The latent heat is then obtained as $x_2 - x_1$. Physically, this implies that when $\beta$ crosses $\beta_c$, we observe a discontinuity in the action of magnitude $x_2 - x_1$.}
\label{Latent Heat Definition}
\end{figure}

The weakening of the first order phase transition was already noted in \cite{Kawai:1992um}, where the "latent heat" of the system, namely the jump in action across the first order phase transition,
was observed to decrease along the line of first order transitions, leaving open the possibility of having a region with vanishing latent heat, a necessary condition for a second order phase transition.
We recall that the "latent heat" $\lambda$ is defined as

\begin{align}
\lambda = x_2 - x_1 \quad \text{ where } \int_{x_1}^{x_2} dx \left( \dfrac{d S_B}{dW_{1 \times 1}^{f}}(x) - \beta_c N_\text{Loops}\right) = 0 \label{eq:Integral Condition}  \\
\dfrac{d S_B}{dW_{1 \times 1}^{f}}(x_2) = \dfrac{d S_B}{dW_{1 \times 1}^{f}}(x_1) = \beta_c N_\text{Loops} \label{eq:Maximum Condition}  \ .
\end{align}

This captures the fact that, if the integral condition is satisfied, the probability of the two actions, $x_1$ and $x_2$, is the same. The value of $\beta$ at which this holds defines the critical coupling $\beta_c$. Note also that this definition is only useful when $x_1\neq x_2$, which is the case for a first order phase transitions. In Fig.~\ref{Latent Heat Definition} we represent schematically the process of determining the critical coupling and, consequently, the latent heat.

\begin{figure}
\centering
\includegraphics[width=0.95 \textwidth]{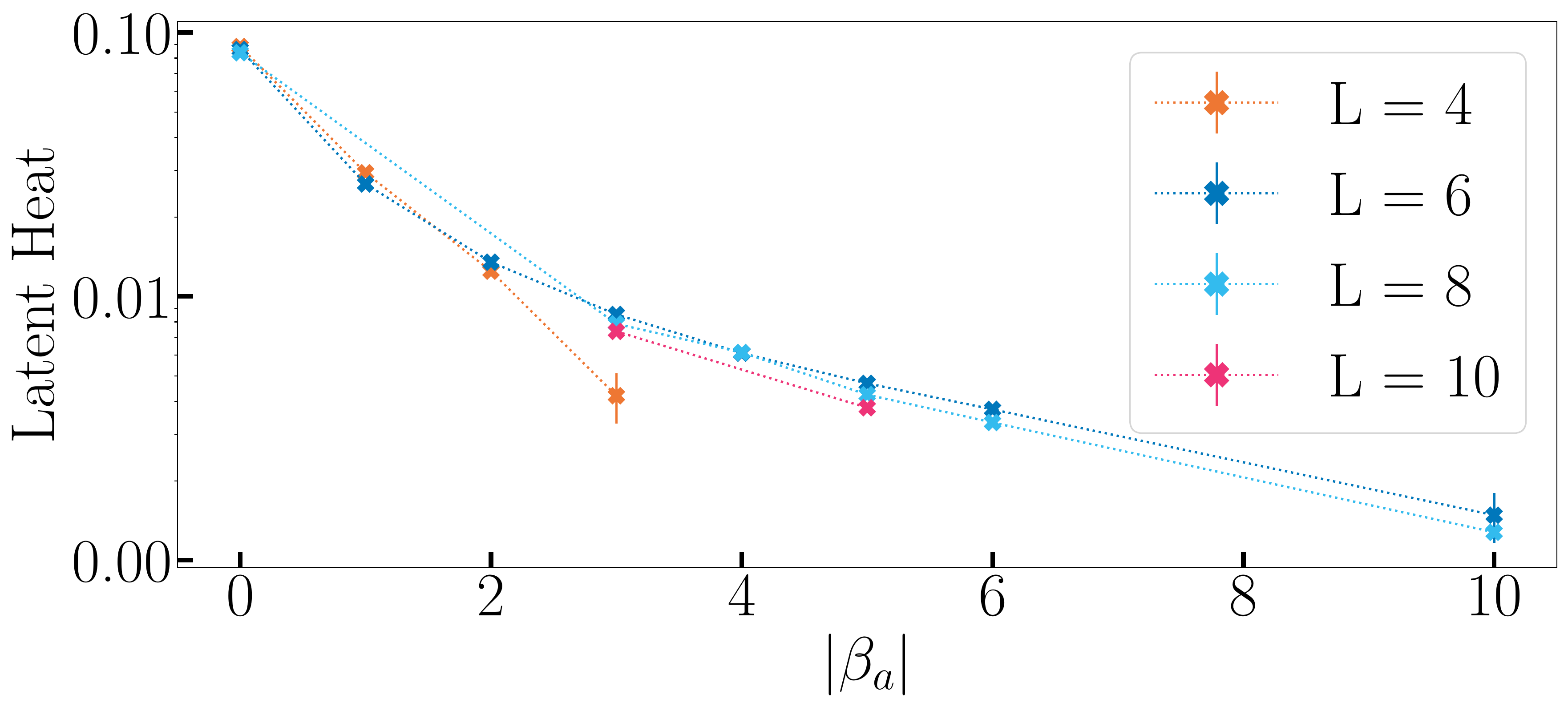}
\caption{Latent heat along the line of first order transition between phases I and III (shown in Fig.~\ref{fig:PhaseSpaceScan}) for different lattice sizes. We observe a relatively strong dependence on the system size, especially for $L=4$, for which the latent heat disappears for $\beta_a<-3$. Note that errors bars are plotted but often smaller than the marker size.}
\label{fig:Latent Heat}
\end{figure}
We reproduce the results of \cite{Kawai:1992um} for $L=4$  and present new ones with $L=6,8,10$ in Fig.~\ref{fig:Latent Heat}. The first phenomenon we observe is the strong dependence on the system size, especially for the smaller $L=4$ lattices. For this size, we do not show more values of $\beta_a$, because there was no unstable region for $\beta_a<-3$. In particular, using the results obtained on the larger lattices, we can rule out the existence of a second order phase transition up to $\beta_a=-10$; this is one of the main results of this work.

Only considering Fig.~\ref{fig:Latent Heat}, we may hope to find a second order transition, at the end of this line of first order phase transitions for some values of $\beta_a$ not too far from $-10$, as was conjectured in \cite{Kawai:1992um} for smaller values of $\beta_a$. We will now argue that this is unlikely to be the case. This will motivate us  to study our second model \eqref{2 Loops Action} to explore a different region in coupling space.

\begin{figure}
\centering
\includegraphics[width=0.5 \textwidth]{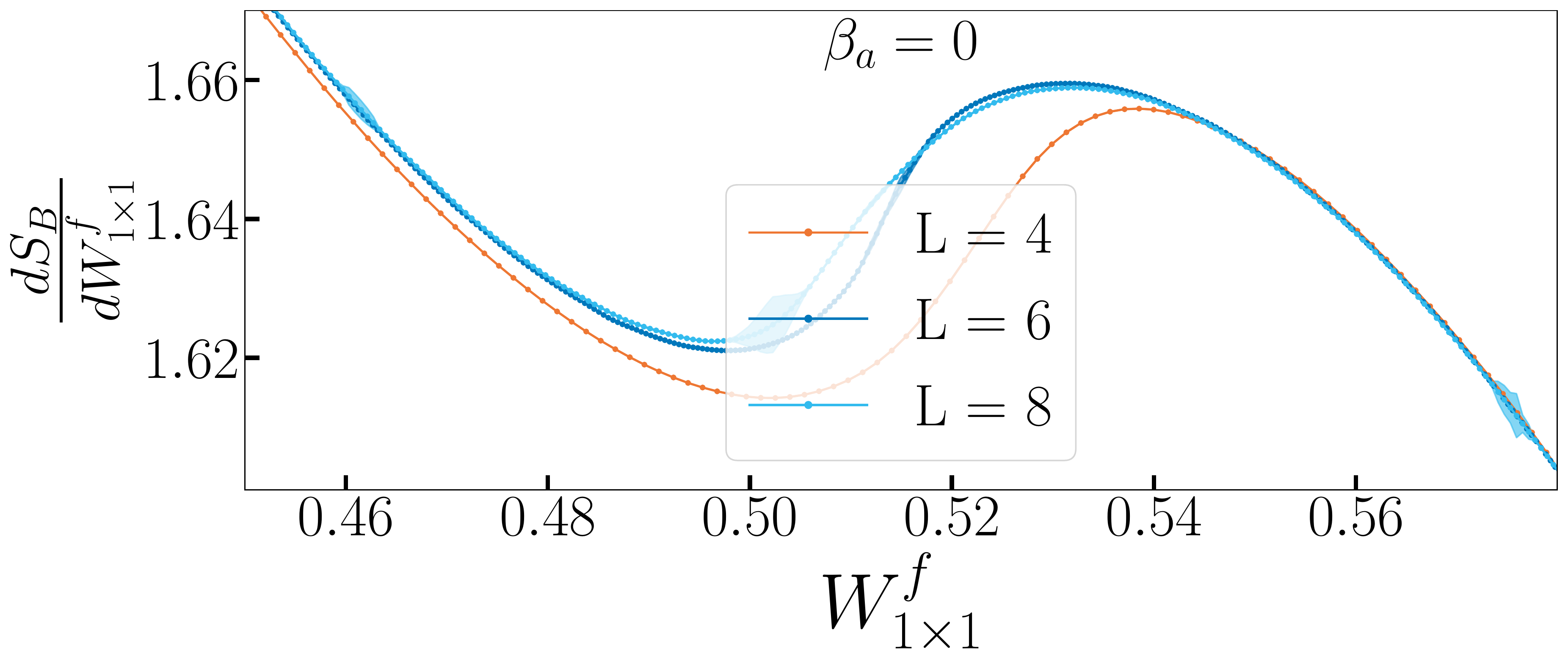}\includegraphics[width=0.5 \textwidth]{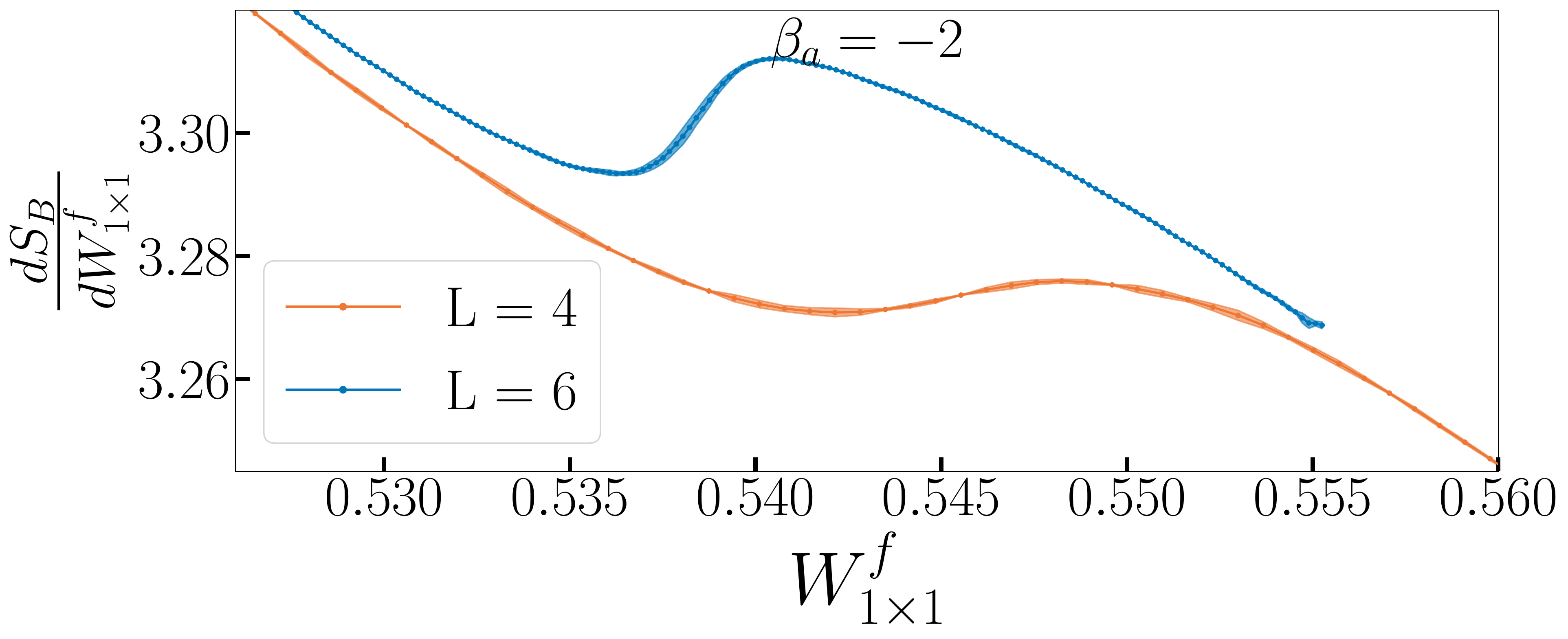}
\includegraphics[width=0.5 \textwidth]{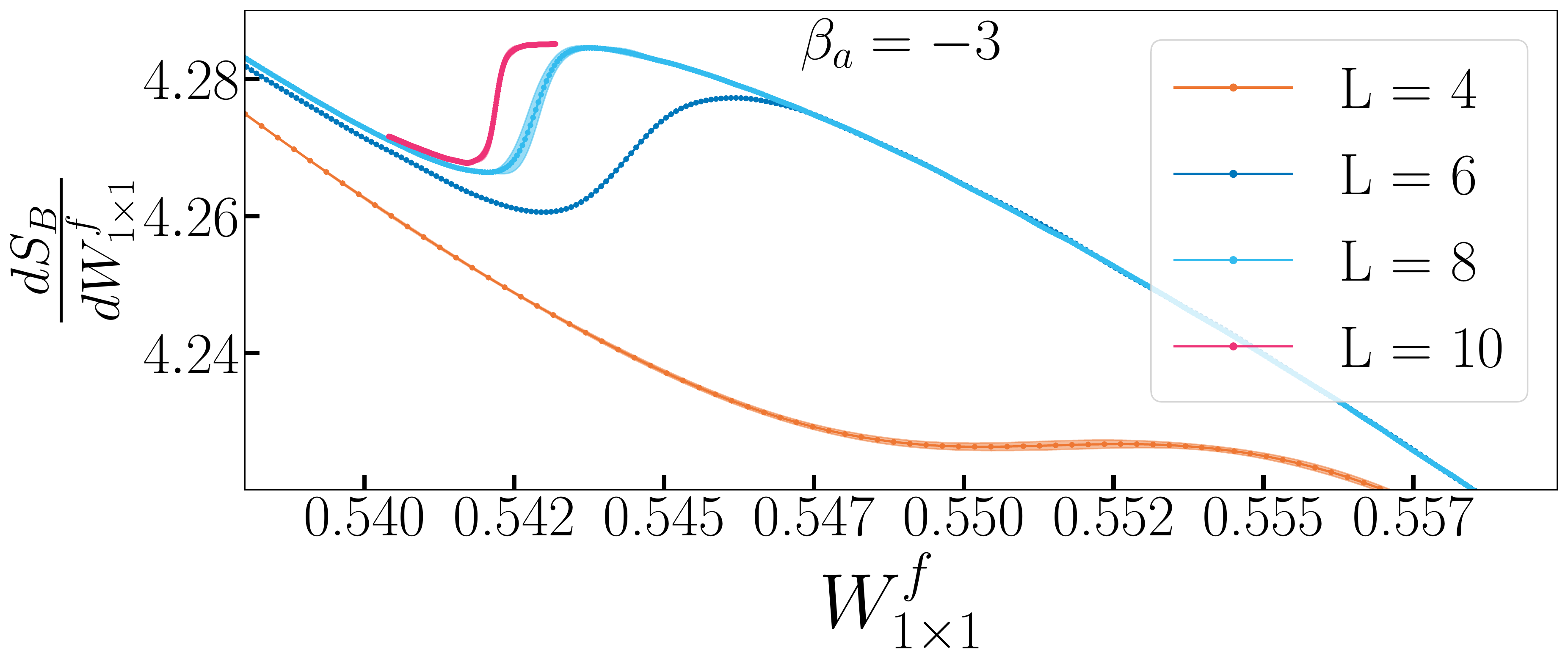}\includegraphics[width=0.5 \textwidth]{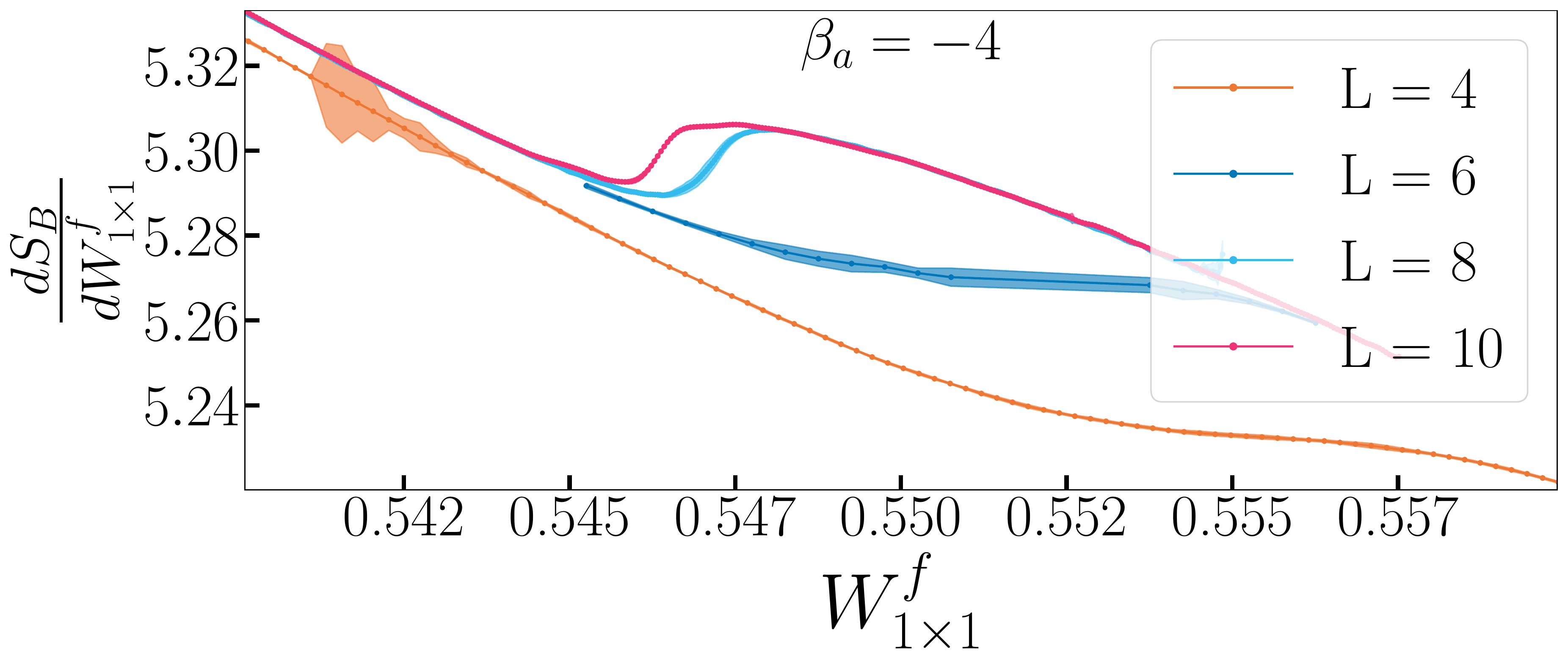}
\includegraphics[width=0.5 \textwidth]{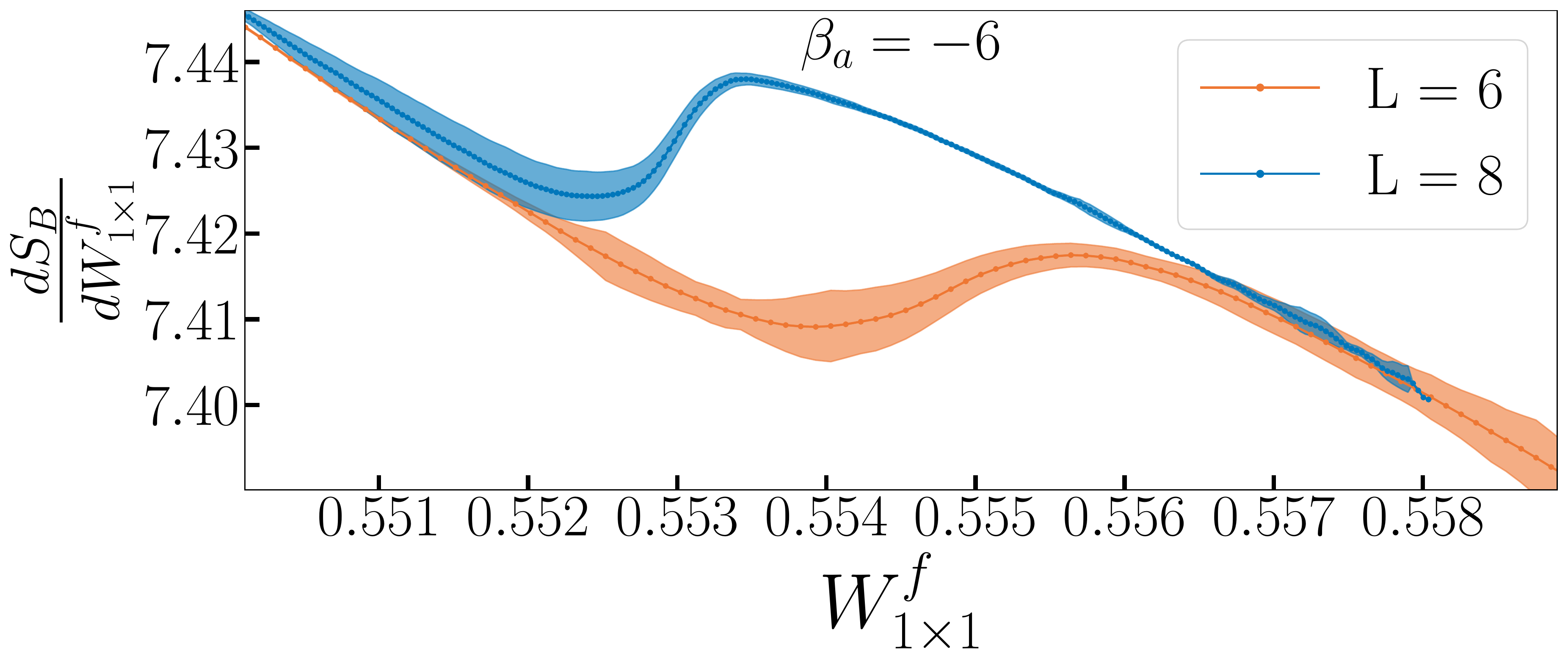}\includegraphics[width=0.5 \textwidth]{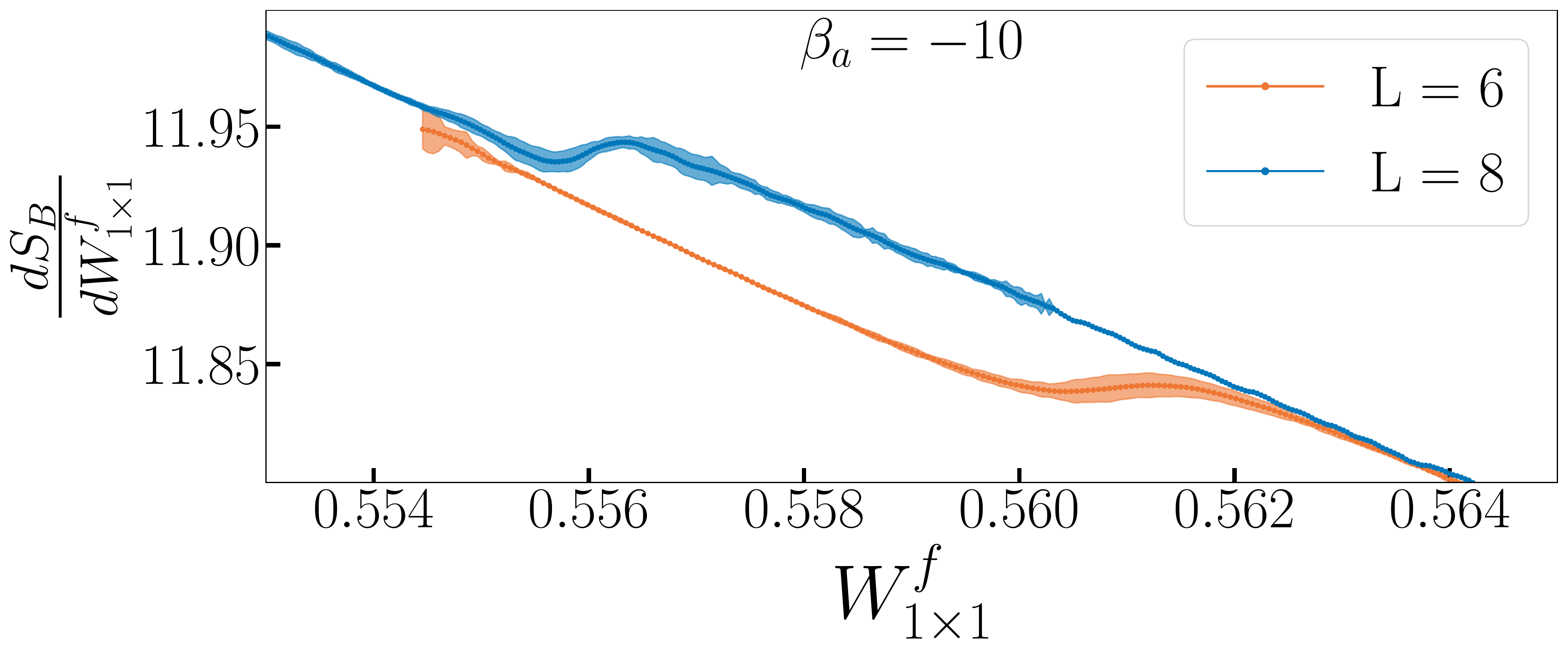}
\caption{First derivative of the entropy with respect to $W^f_{1\times 1}$, as a function of $ W^f_{1\times 1} $, for different lattice sizes and $\beta_a$. The value of $\beta_a$ has a large impact on the dependence on the lattice size.  The colored region around the markers is an estimate of the error based on independent measurements.
}
\label{fig:First Derivative Adjoint}
\end{figure}

To study in details the nature and strength of the phase transition, we now resort to the method introduced in Sec.~\ref{sec:mircocanonical} and perform simulations in the modified ensemble \eqref{eq:new_ensemble}. This allows us to obtain the results for the derivative of the microcanonical entropy presented in Fig.~\ref{fig:First Derivative Adjoint}. As already explained in the previous section, the existence of a convex region signals the existence of a first order transition as it is associated with the existence of two distinct minima of the free energy and thus two distinct phases. It is worth stressing again that these results are out of reach of conventional "canonical" simulations, as this unstable region corresponds to configurations which have vanishingly small
probability and which thus, for all practical purposes, do not contribute to the path integral and are, as a result, not sampled by conventional Monte-Carlo.

The results presented in Fig.~\ref{fig:First Derivative Adjoint} follow a clear pattern. While the latent heat, which can be read from the size of the unstable region on these plots, does indeed decrease with $\beta_a$, the measurements for all $\beta_a$ show a sharp increase in the intensity of the unstable region; the second derivative of the entropy (namely the slopes of Fig.~\ref{fig:First Derivative Adjoint}) sharply increase as the infinite volume limit is approached (we recall that this is the intensive entropy).  We illustrate this behaviour for $\beta_a=-3$ in Fig.~\ref{fig:Second Derivative Adjoint}. This is incompatible with the existence of second order phase transition, as  the second derivative of the microcanonical entropy needs to vanish. Moreover, it usually approaches zero from below.

This consistent increase in the second derivative of the entropy as the infinite volume limit is approached leads us to believe that a second order order phase transition is unlikely to exist in a region of parameter space close to $\beta_a=-10$. Together with the increased dependence on the system size observed with the decrease of $\beta_a$, see Fig.~\ref{fig:First Derivative Adjoint}, it provides compelling arguments in favor of exploring a different region of parameter space; this is what we do in the next section by considering a coupling to square loops of size two.

\subsection{Variable Size Wilson Loops}
\label{sec:res2b2}

We now move on to study a different part of coupling space. As we argued previously, because of the relation between the trace in the fundamental representation and the trace in the adjoint representation, we expect $\beta_f$ and $\beta_a$ to be correlated.
To have more freedom, we now study the model described by the action \eqref{2 Loops Action} with two independent couplings $\beta_1$, to fundamental plaquettes, and $\beta_2$, to two by two Wilson loops in the fundamental representation.

As before, we start by scanning the parameter space for a small lattice size ($L=4$) to identify the interesting regions of the parameter space. Once again, we do not show the region for $\beta_1 < 0$, due to the existence of a reflection formula
\[
\left\langle W^f_{1\times 1} \right\rangle _{\beta_{1},\beta_{2}}+\left\langle W^f_{1\times 1}\right\rangle _{-\beta_{1},\beta_{2}}=2,\qquad \qquad \left\langle W^f_{2\times 2}\right\rangle _{\beta_{1},\beta_{2}}=\left\langle W^f_{2\times 2}\right\rangle _{-\beta_{1},\beta_{2}} \ . \label{eq:transformation2loops}
\]
which follows from the symmetry \eqref{eq:sym} (with $\beta_1$ playing the role of $\beta_f$).

\begin{figure}
\centering
\phantom{a}

\includegraphics[width=1.02 \textwidth]{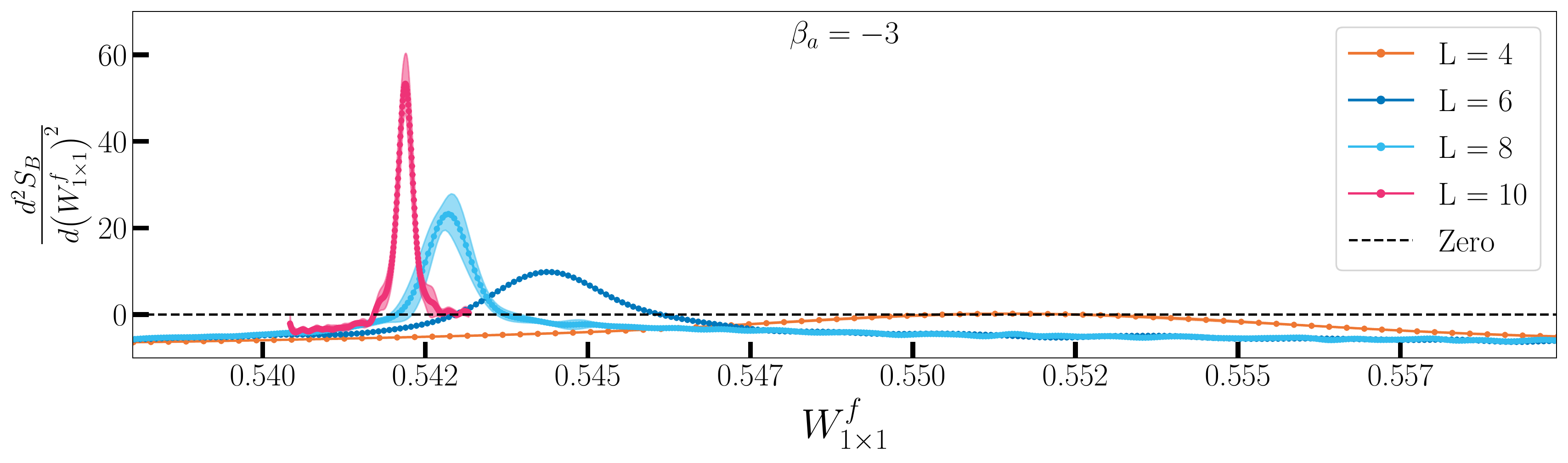}
\caption{Second derivative of the entropy with respect to $W^f_{1\times 1}$, as a function of $ W^f_{1\times 1} $, for $\beta_a = -3$ and different lattice sizes. The dependence on the lattice size becomes even more pronounced than in the case of the first derivative. The colored region around the markers is an estimate of the error based on independent measurements.}
\label{fig:Second Derivative Adjoint}
\end{figure}

\begin{figure}
\centering
\includegraphics[width=0.45 \textwidth]{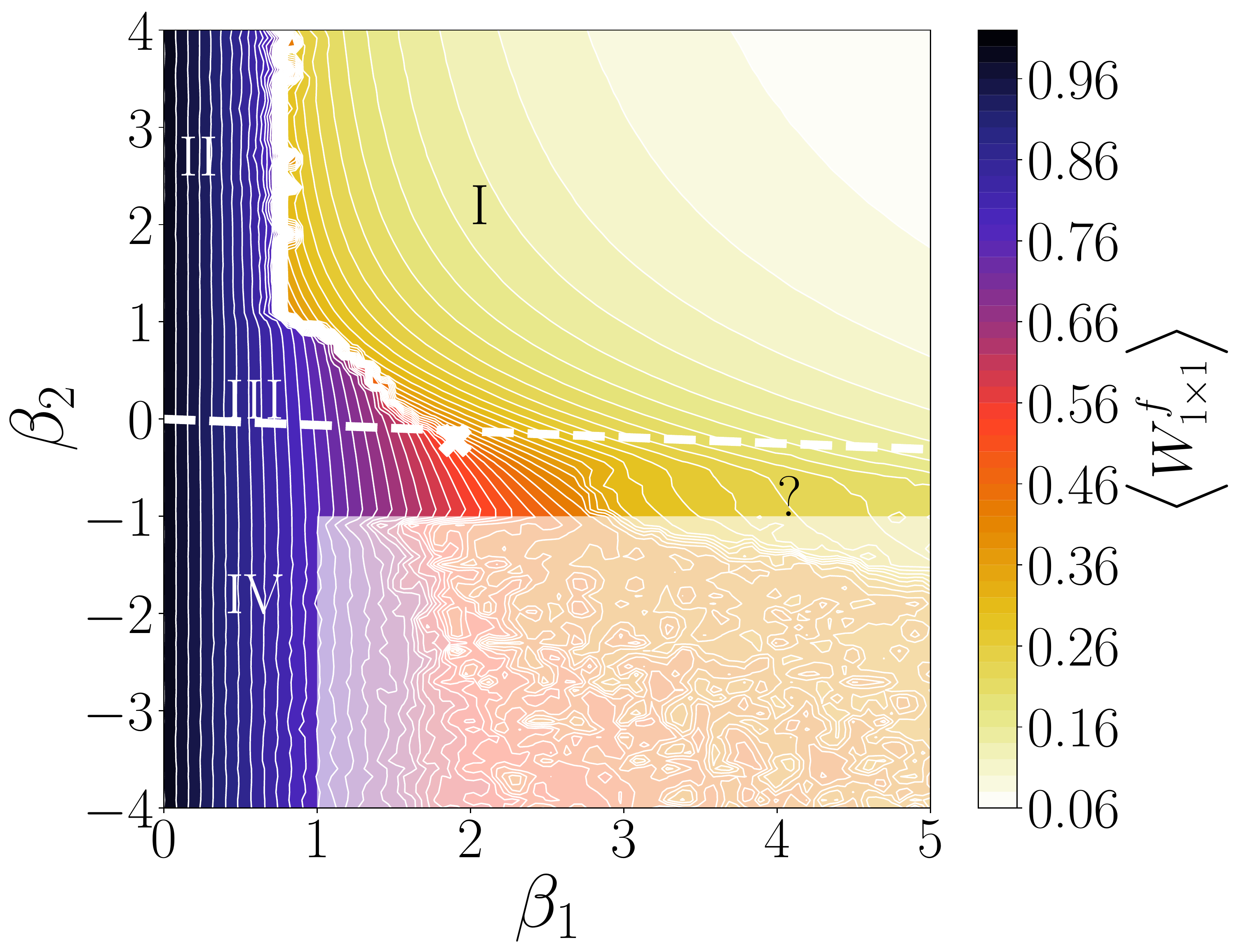}\includegraphics[width=0.45 \textwidth]{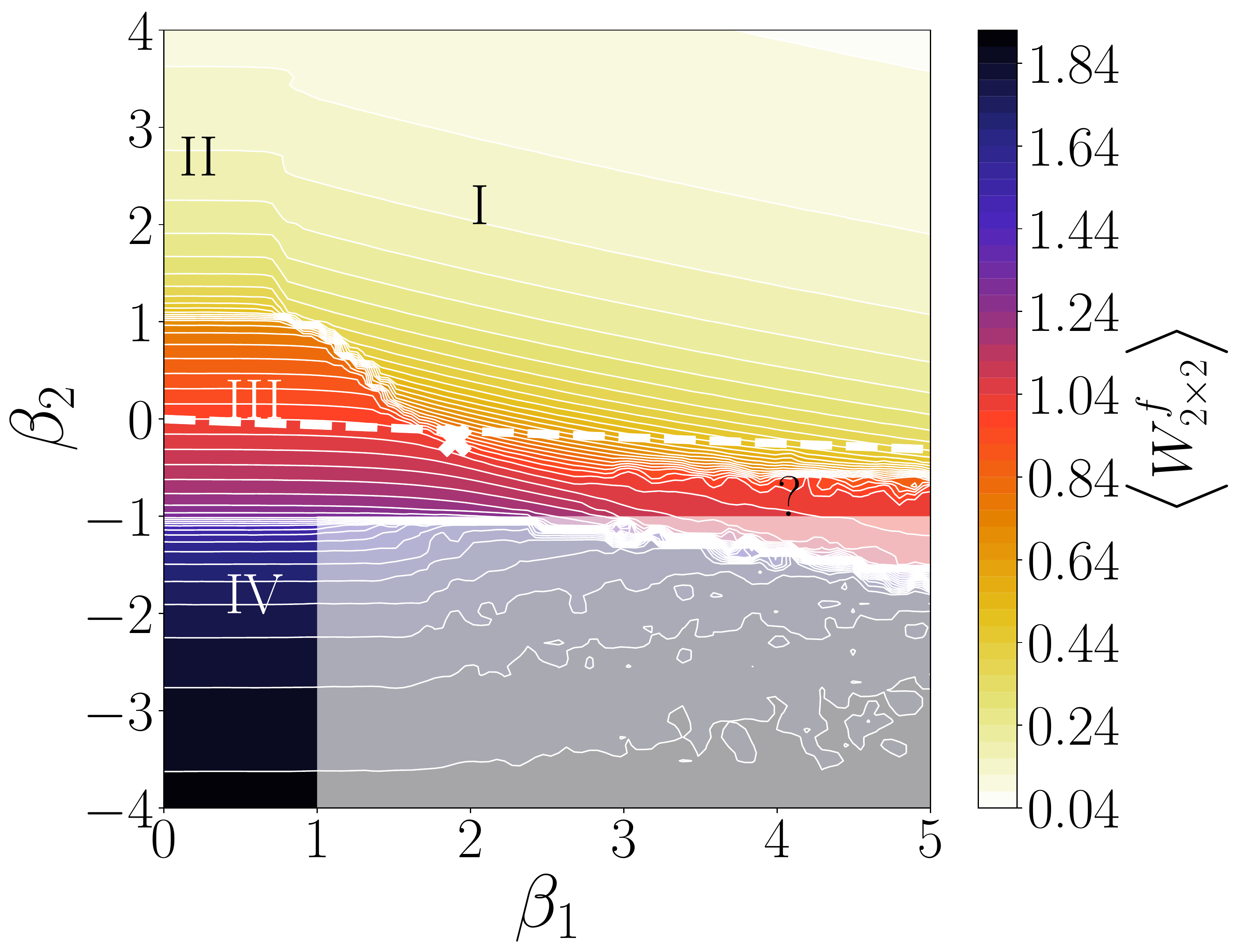}
\caption[Phase Space Scan 2 Loops]{Survey of the parameter space, for $L=4$. The image on the \textbf{left} hand-side shows $\left\langle  W^f_{1\times 1}\right\rangle _{\beta_{1},\beta_{2}}$ and the one on the \textbf{right} hand-side $\left\langle  W^f_{2\times 2}\right\rangle _{\beta_{1},\beta_{2}}$. The white cross marks the point we selected for subsequent analysis. We clearly identified four different phases, labeled by  roman-numerals. The thin white lines are level curves, therefore the closer they are, the quicker the change in value. Hence, we can identify discontinuities in the values sampled when several white lines come together. In the bottom right corner (washed out region), we observe frustration. The dashed white line distinguishes regions with different signs of the coupling, for the naive continuum limit ($\nicefrac{1}{g^2} < 0$ below the line, $\nicefrac{1}{g^2} > 0$ above the line). This figure also seems to  leave open the possibility of the existence of a fifth phase, denoted by a question mark (?).}
\label{Scan 2Loops}
\end{figure}

The results, shown in Fig.~\ref{Scan 2Loops}, suggest the existence of four different phases, identified by roman numerals.
When either $\beta_f$ or $\beta_a$ is zero, we are working with only one type of fundamental loops and expect the usual confinement/deconfinement phase transition. Taking into account the symmetry \eqref{eq:transformation2loops}, it explains the phase transition between phases III and II, III and IV and III and I. The phase transition between phase II and phase I  is also not a surprise and is  similar to the phase transition between phase II and I in the previous model. The theory with only two by two loops, for a given lattice size, has more vacua than the theory containing only plaquettes. Indeed, any configuration which is a minimum of the plaquette action will also minimize the 2 by 2  loops action. The converse is not true. This can be seen by taking a configuration that minimizes the plaquette action and multiply a line of links by minus -1. The resulting configuration is still a minimum of the 2 by 2 Wilson loops action (every plaquette contains two reflected links) but generically not of the plaquette action. This also means that for very large values of $\beta_2$ the configurations which dominates the statistical sum may take arbitrary values of $\text{Tr}\left(\square_{1\times 1}^f\right) $, leading to a vanishing expectation value. Thus $W_{1\times 1}^f\approx 1$ for large $\frac{\beta_2}{\beta_1}>>1$ and this is what we observe in Fig.~\ref{Scan 2Loops}. As phase I supports non-vanishing averages of  $\text{Tr}\left(\square_{1\times 1}^f\right) $, they are two distinct phases separated by a phase transition.

The shaded region on the bottom right corner of both color maps in Fig.~\ref{Scan 2Loops} shows a region we could not satisfactorily  sample as we observed frustration (long plateaus in the action, along the Markov chain, followed by jumps several times bigger than the fluctuations observed on those plateaus). This  particularly affects the measurements of $\left\langle W^f_{1\times 1}\right\rangle$ and, for this reason, we ignored this region in the analysis of the distinct phases we carry later on. Note also that right at the boundary of this region, the data may suggest the existence of yet an extra phase, denoted by a question mark in  Fig.~\ref{Scan 2Loops}. The difficulty to reliably sample this region of parameter space did not allow us to study this question in any detail and is thus left for future work.

With the relevant regions identified, we can study the nature of the phase transitions. The phase transitions between II and III, and III and IV, are necessarily first order, as they are the equivalent to the ones present already in the model with only the fundamental action. We also measured strong first order phase transitions between phase I and II. The remaining and promising possibility to uncover a second order phase transition is between phase I and III; we highlighted in Fig~\ref{Scan 2Loops} the region where we will focus our effort by a white cross.

\begin{figure}
\centering
\includegraphics[width=0.95 \textwidth]{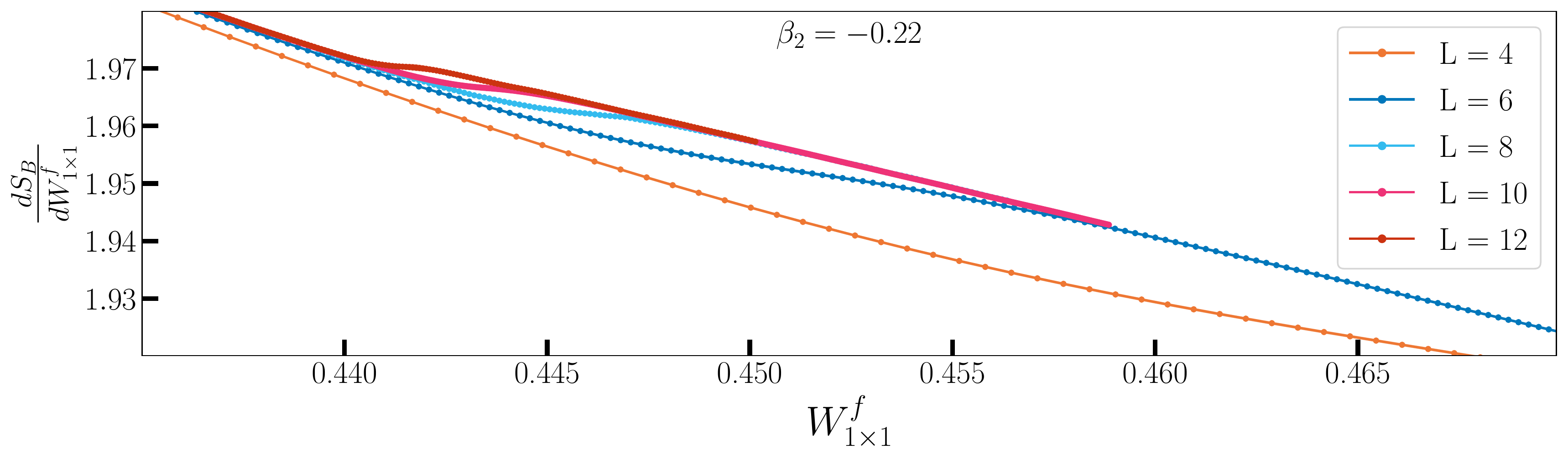}
\includegraphics[width=0.95 \textwidth]{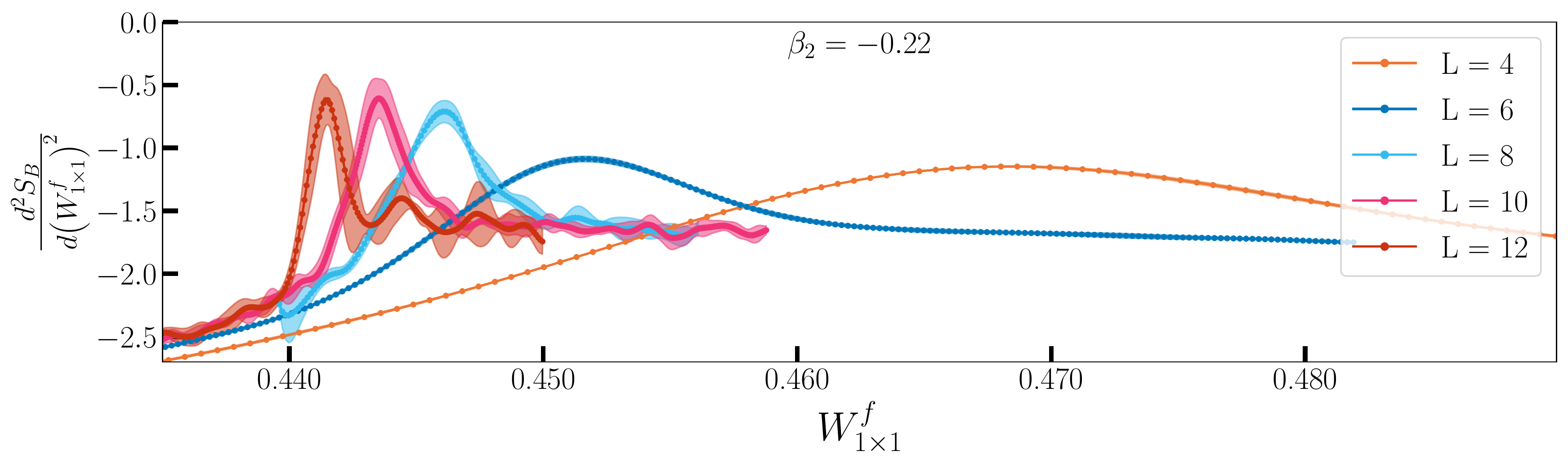}
\caption{First (\textbf{top}) and second derivative (\textbf{bottom}) of the entropy as a function of $W^f_{1\times 1}$, for different lattice sizes and $\beta_2=-0.22$. The colored region around the markers is an estimate of the error based on independent measurements. }\label{First and Second derivative 2Loops}
\end{figure}

We also measure the first and second derivatives of the entropy. As expected from the color plot Fig.~\ref{Scan 2Loops}, we obtain first order phase transition for small beta values. As in the previous model, small lattices suggest this phase transition disappears at some values of $\beta_2$. We recall that there, we found this to be an artifact of small lattices and that the first order transition always seemed to reappear for larger lattice, with no real signs of weakening.
This is in complete contrast with what we obtain for this model.
We show for instance results at $\beta_2=-0.22$ in Fig.~\ref{First and Second derivative 2Loops}.
Contrary to what was observed in the previous case in Fig.~\ref{fig:First Derivative Adjoint} and Fig.~\ref{fig:Second Derivative Adjoint}, we see no clear signs of the emergence of a  phase transition  as the volume increases.

To try to understand better how robust the disappearance of the first order phase transition is against the infinite volume limit and if we can see hints of the emergence of second order criticality, we study in more detail the maximum of the second derivative of the entropy. As already mentioned, a positive second derivative signals the existence of an unstable region and thus of a first order phase transition. Only non-positive values for the second derivative will corresponds to either a smooth crossover or a continuous phase transition of the second type discussed in section \ref{sec:prediction}.

\begin{figure}
    \centering
    \includegraphics{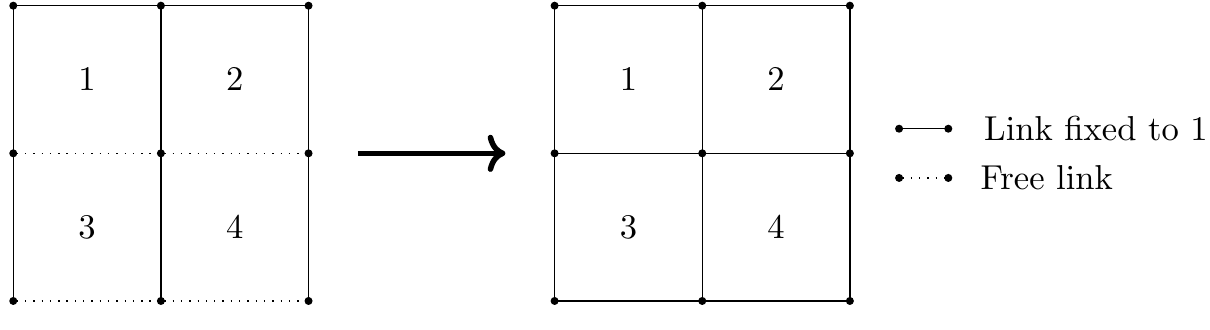}
    \caption{Impossibility of minimizing, at the same time, the contributions from the loops of size one and the contributions from the loops of size 2 when $\beta_1 > 0$ and $\beta_2 < 0$. Starting from a gauge fixed configurations with all the links set to $\mathbb{I}$ except for one per plaquette, we minimize the action for the loop of size one by imposing $\text{Tr}\left(\square_{1\times 1}^f\right) = 1 $, resulting in the configuration in the right hand-side. This configuration is clearly incompatible with the condition $\text{Tr}\left(\square_{2\times 2}^f\right) = -1$ required to minimize the action for loops of size two on its own.}
    \label{fig:impossibility_of_minimizing_both_actions_v2}
\end{figure}

The frustration appearing in the right bottom corner of Fig.~\ref{Scan 2Loops} can be understood as the impossibility of minimizing, at the same time, the contributions for the action coming from plaqutetes of size one and two, when $\beta_1 > 0$ and $\beta_2 < 0$.
To understand this, we can look at a 2-dimensional slice of the 5-dimensional lattice, and try to find a configuration that minimizes both contributions. We start by working in a gauge where all links in a plaquette are fixed to the identity $\mathbb{I}$ except for one per plaquette, as depicted in the left hand-side of Fig.~\ref{fig:impossibility_of_minimizing_both_actions_v2}, see also Ref.~\cite{Gattringer:2010zz}. We now minimize the action for the loops of size one with $\beta_1 > 0$. In this case, we want to find configurations such that $\text{Tr}\left(\square_{1\times 1}^f\right) = 1 $, for all loops of size one.  Consider  loop 1 in Fig.~\ref{fig:impossibility_of_minimizing_both_actions_v2}. Three out of the four links are already set to $\mathbb{I}$, such that, in order for the trace over this loop to be 1 the  bottom gauge link must be also $\mathbb{I}$. The same argument works for loop 2. This then implies the same for loops 3 and 4, resulting in the configuration shown in the right hand-side of Fig.~\ref{fig:impossibility_of_minimizing_both_actions_v2} with all the links set to $\mathbb{I}$.
It is then clear that it is now impossible to minimize the action for the loops of size two (with $\beta_2 < 0$). Indeed,  this would require $\text{Tr}\left(\square_{2\times 2}^f\right) = -1$. We stress that this obstruction to minimize both actions at the same time is extensive and is not a finite volume effect.

\begin{figure}
\centering%\begin{mdframed}[backgroundcolor=blue!50,linecolor=blue!50]

\includegraphics[width=0.95 \textwidth]{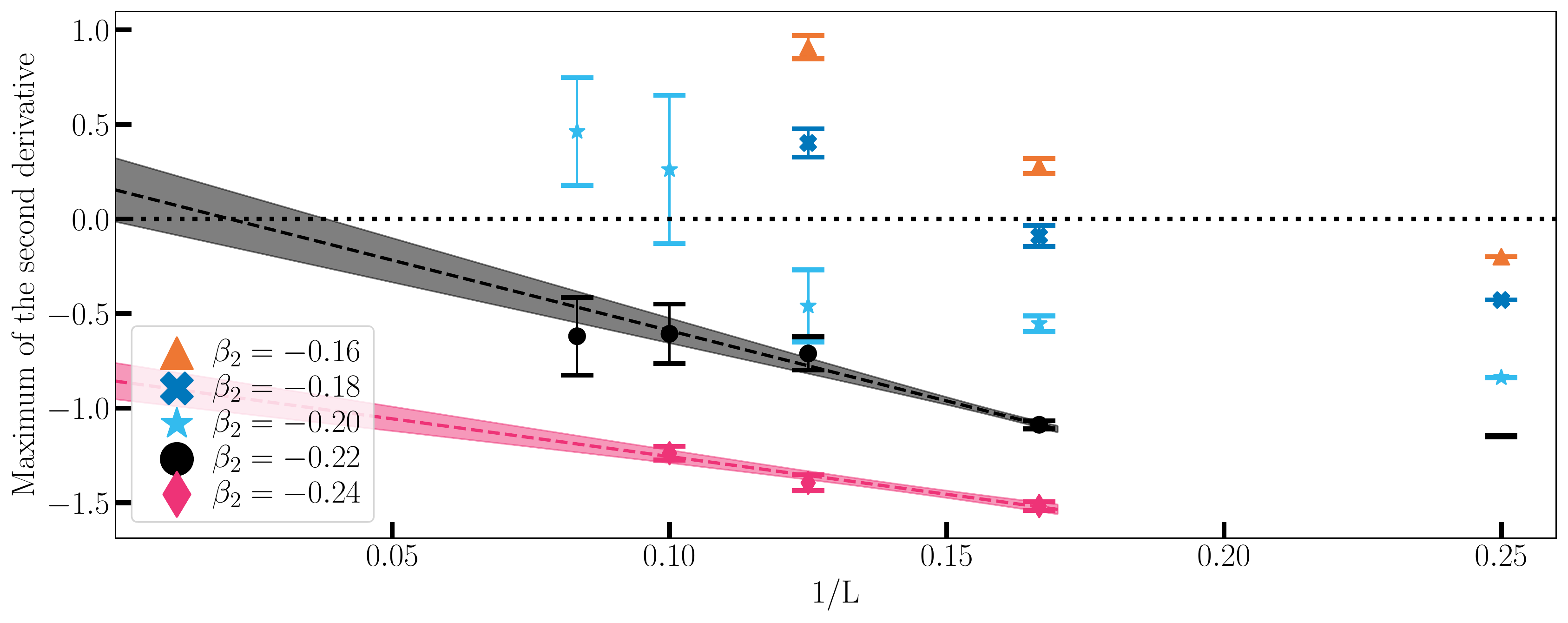}
\caption{On the \textbf{top} panel, we show the dependence of the maximum of the second derivative of the entropy as a function of $\nicefrac{1}{L}$, for different $\beta_2$ (colors).  The error bars are obtained by generating histograms from independent simulations. For $\beta_2 = -0.22$ and $\beta_2 = -0.24$, we show a fit (dashed line) and the respective uncertainty (shaded region), to show that the decreasing of the second derivative with $\beta_2$, may result in a second order phase transition. We want to emphasize that the extrapolations are more qualitative than quantitative and that larger lattices are required to perform a controlled infinite volume limit. }
\label{Maximum Secon Derivative}
%\end{mdframed}
\end{figure}

We summarize in Fig.~\ref{Maximum Secon Derivative} the results for the maximum of the second derivative of the entropy for different values of $\beta_2$ (around -0.20) as a function of $1/L$. For larger $\beta_2$ values, namely $\beta_2=-0.16,-0.18,-0.20$ , we see that a first order phase transition is recovered. Something more interesting happens around $\beta=-0.22$ (black dots in Fig.~\ref{Maximum Secon Derivative}), where we focused our efforts, generating better data sets (this point is signaled by the withe cross in Fig.~\ref{Scan 2Loops}; the first and second derivatives of the entropy for this point are represented in Fig.~\ref{First and Second derivative 2Loops}). The maximum of the second derivative of the entropy is negative  for all our lattice sizes. More importantly, the volume dependence seems to be weak enough for this behavior to survive the infinite volume limit. Of course, because of the smallness of our lattices, we, unfortunately, cannot take any reliable infinite volume limit. This said, the finite volume corrections to the data for $\beta_2=-0.22$, excluding the $L=4$, are reasonably well modeled by $1/L$ corrections. As a result, we also show a tentative linear extrapolation to the infinite volume limit together with our data points. We see that, to the extent this extrapolation can be trusted, the data is consistent with a second order phase transition or a very weak first order phase transition.

What makes this region of coupling space particularly interesting is that, independently of volume, we observe a consistent decrease in the value of the second derivative when $\beta_2$ decreases. We further illustrate this by showing points (pink diamonds) for yet a more negative value of $\beta_2$, namely $\beta_2=-0.24$. Note that these data become increasingly harder to collect the more negative $\beta_2$ becomes, as the apparent disappearance of the critical behavior requires more statistics to be satisfactorily resolved.

Note also that we plot together with the data points a (rather poor) linear extrapolation to the infinite volume limit, more to guide the eyes of the reader than to make any quantitative predictions. All the data points (for $\beta_2 = -0.24$) are smaller than for $\beta_2=-0.22$ and the extrapolation to the infinite volume limit is consistent with being either smaller or equal to the $\beta_2=-0.22$ one. This means that, even if for instance the case $\beta_2=-0.22$ turns out to be a very weak first order phase transition, it seems plausible that further decreasing $\beta_2$ will lead to a maximum of the second derivative that remains non-positive in the infinite volume limit and thus correspond to a second order phase transition.

The questions  of whether this fixed point actually exists, what is its precise location and whether it is part of a line of second order phase transitions are unfortunately impossible to settle with any confidence given our data (and lattice size limitations) and are thus left for future work.

%\begin{figure}
%\centering
%\includegraphics[width=0.95 \textwidth]{MaximumSecondDerivative2.pdf}
%\caption{ On the \textbf{top} panel, we show the dependence of the maximum of the second derivative of the entropy as a function of $\nicefrac{1}{L}$, for different $\beta_2$ (colors).  The error bars are obtained by generating histograms from independent simulations. For $\beta_2 = -0.22$ and $\beta_2 = -0.24$, we show a fit (dashed line) and the respective uncertainty (shaded region), to show that the decreasing of the second derivative with $\beta_2$, may result in a second order phase transition. We want to emphasize that the extrapolations are more qualitative than quantitative and that larger lattices are required to perform a controlled infinite volume limit. }
%\label{Maximum Secon Derivative}
%\end{figure}

\section{Conclusions}

This work aimed to look again for putative non-trivial UV fixed points in the coupling space of $SU(2)$ Yang-Mills in five dimensions, as suggested by the $\epsilon$ expansion in $4+\epsilon$ dimension.
To this end, we introduced two different lattice models in Sec.~\ref{subsec:models}. Then, thanks to the knowledge gained in \cite{Kawai:1992um}, we were able to anticipate the problem of having to be able to distinguish between weak first order phase transitions with small latent heats and true second order phase transitions. As a result, we imported from the condensed matter literature \cite{velazquezExtendedCanonicalMonte2016} and presented in Sec.~\ref{sec:mircocanonical} an improved method based sampling the system of interest in a modified ensemble. In particular, we argued that this method makes the measurement of the microcanonical entropy over phase transitions possible and that it is a helpful tool to study the order of such transitions.

With this at hand, we moved on to present the results obtained for our two different models. We started in Sec.~\ref{subsec:resfa} by discussing the model with fundamental plaquettes and adjoint plaquettes, see Eq.~\eqref{Fundamental + Adjoint Action}. Our main new result, in this case, is that despite the legitimate hopes raised by the results of \cite{Kawai:1992um}, thanks to our new data obtained on bigger lattices and mostly thanks to our refined analysis performed with the methods of Sec.~\ref{sec:mircocanonical}, a second order phase transition is excluded for adjoint couplings up to $\beta_a=-10$ and unlikely to exist for yet smaller $\beta_a$. This negative result encouraged us to move on to a second model made of plaquettes and two by two square Wilson loops, see Eq.\eqref{2 Loops Action}, whose results are presented in Sec.~\ref{sec:res2b2}. In this case, we were able to observe a disappearance of a line of first order transition, a disappearance that seems likely to survive the infinite volume limit. Our data at the end of this line of first order transitions are compatible with the existence of a second order phase transition, even though they are not good enough to assert its existence and determine its potential precise location.
Note also that, as seen in Fig.~\ref{Scan 2Loops}, since the transition between phase I and III is weakly dependent on $\beta_1$, an efficient strategy would be to repeat our analysis for fixed $\beta_1$.

In a nutshell, more powerful computers and newer algorithms have helped us make progress from the work of \cite{Kawai:1992um} by excluding the existence of a second order phase transition in the region of parameter space located there. It also helped us to put forward a different region in coupling space which is a good candidate to look for a second order critical point. It has unfortunately not been sufficient to locate such a fixed point with enough confidence. We hope that this time less than a generation will pass before this region of the coupling space is further investigated.

\section*{Acknowledgments}

We are particularly grateful to P. de Forcrand for several useful comments and suggestions. We would also like to thank D.~Clarke, G.~Cuomo, E.~Grossi, M. Luscher and S. Rychkov for interesting comments and discussions.
AF is supported by the Swiss National Foundation and the U.S.Department of Energy, Office of Science, Office of Nuclear Physics, grants Nos. DE-FG88ER40388.
JP is supported by the Simons Foundation grant 488649 (Simons Collaboration on the Nonperturbative Bootstrap) and by the Swiss National Science Foundation
through the project 200021-169132 and through the National Centre of Competence in Research SwissMAP. JVL and JM   acknowledge support   by   the   Portuguese   Foundation   for   Science   and   Technology   through   Strategic Funding   No. UIDB/04650/2020 and   Project   No.   POCI-01-0145-FEDER-028887.
 The computations in this paper were run on the EPFL SCITAS cluster, the  Grid FEUP   Avalanche cluster and CIRRUS A through the project CPCA/A0/7287/2020.

\appendix

\section{Fundamental and adjoint representations of SU(2)}
\label{app:SU2RepTheory}
There is a nice relation between the trace of an SU(2) element in the fundamental and in the adjoint representation, which we will derive in this section. We can represent a generic element, $g$, of $SU(2) = \left\{ M \in M_{2\times2}(\mathbb{C}) | \det(M) = 1 \right\}$, in the fundamental representation as
\[
g = \left(\begin{array}{cc}\alpha & -\bar{\beta} \\ \beta & \bar{\alpha}\end{array}\right).
\]
The adjoint representation is given by the derivative, at the origin, of the conjugacy map, $\Psi:SU(2)\times SU(2)\to SU(2)$ given by $\Psi(g,h)= ghg^{-1}$, or alternatively the map  $\Psi_g:SU(2)\to SU(2)$ given by $\Psi_g(h)= ghg^{-1}$. By representing $h$ in the Lie algebra by the coordinates \{$\omega_i$\}, the derivative of $\Psi_g$ at the origin becomes
\[
\begin{aligned}
  (\Psi_g)_{*}(e)& = \left.\frac{d \Psi_{g}(h)}{d \omega^{i}}\right|_{\omega_i=0}\\
&=gu_jg^{-1}\\
\end{aligned}
\]
We now study how this object acts on a generic element of the Lie algebra, $x$, and define the adjoint of $g$, $Ad(g)$, as the map such that
\[
Ad(g) x =  (\Psi_g)_{*}(e) (x),
\]
yielding
\[
Ad(g) = \operatorname{Ad}\left(\left(\begin{array}{cc}\alpha & -\bar{\beta} \\ \beta & \bar{\alpha}\end{array}\right)\right)= \left(\begin{array}{ccc}\Re\left(\alpha^{2}-\beta^{2}\right) & -\Im\left(\alpha^{2}-\beta^{2}\right) & \beta \bar{\alpha}+\alpha \bar{\beta} \\ \Im\left(\alpha^{2}+\beta^{2}\right) & \Re\left(\alpha^{2}+\beta^{2}\right) & i(\beta \bar{\alpha}-\alpha \bar{\beta}) \\ -(\alpha \beta+\bar{\alpha} \bar{\beta}) & -i(\alpha \beta-\bar{\alpha} \bar{\beta}) & \alpha \bar{\alpha}-\beta \bar{\beta}\end{array}\right).
\]
Taking the trace of the fundamental representation yields
\[
\operatorname{Tr} \left(\left(\begin{array}{cc}\alpha & -\bar{\beta} \\ \beta & \bar{\alpha}\end{array}\right)\right) = 2 \Re{\alpha},
\]
while the trace of the adjoint representation becomes
\[
\operatorname{Tr} \left(\operatorname{Ad} \left(\left(\begin{array}{cc}\alpha & -\bar{\beta} \\ \beta & \bar{\alpha}\end{array}\right)\right)\right) = (2 \Re{\alpha})^2 - 1.
\]
We finally conclude that
\[
\operatorname{Tr} (Ad(g)) = \operatorname{Tr} (g)^2 - 1.
\]
\section{Monte-Carlo algorithms}
\label{App:MonteCarloAlgorithms}

In this appendix, we discuss the algorithms used to generate our lattice configurations. In the case of the adjoint model \eqref{Fundamental + Adjoint Action}, we used as a base algorithm the generalized multi-hit Metropolis ("Independent Multiple Try Metropolis") of \cite{Martino_2018} with $5$ tries per steps. In the case of the $2$-loops model  \eqref{2 Loops Action}, we designed an "almost heat-bath" algorithm (partial heat-bath with an "accept-reject" step) which we will describe below.

In both cases, we also used some parallel tempering \cite{doi:10.1142/S0129183196000272} to improve thermalization. To further decreased the autocorrelation time, we also performed some overrelaxation steps on the fundamental part of the action, corrected by an "accept-reject" step weighted by the second part of the action (i.e. either adjoint or 2-loops contribution).

\subsubsection*{Heat Bath}\label{Poor mans heat bath}

When all loops are in the fundamental representation (as is the case for \eqref{2 Loops Action}), we developed a mixed heat-bath metropolis algorithm, with a rejection probability suppressed with the system size, allowing us to have the efficiency of the heat bath while retaining our modified ensemble.

 Let $U$ be the lattice link we will to update and let us denote by $\sqsupset^\mathfrak{f}_{\ell}$ the ordered product of the links around a square loop of side $\ell$, except for $U$, such that \[
\square^f_{\ell \times \ell} = U \sqsupset^f_{\ell}.
\]
Recall that there are several ordered products of links that, when $U$ is added, form a closed loop. For our purposes, it is irrelevant to distinguish them and we will redefine this symbol to mean the sum over all such links
\[
\sqsupset^f_{\ell}\to \sum_{\sqsupset^f_{\ell}} \sqsupset^f_{\ell}.
\]
This sum is not an element of $SU(2)$. However, $SU(2)$ has the nice property that the sum of its elements is proportional to an element of $SU(2)$, such that
\[
\sqsupset^f_{\ell}  = c^f_{\ell} A^f_{\ell},
\]where $A^f_{\ell}$ is an element of $SU(2)$ and $c^f_{\ell} \in \mathbb{R}$.
The procedure goes as follows
\begin{enumerate}
    \item Define
    \begin{align}
     \alpha_1 &= \omega_{1} + 2 \omega_{2} \left( W^{f \prime}_{1 \times 1} + \dfrac{2 (D - 1)}{N_{\text{Loops}}} \right) \\
     \alpha_2 &= \frac{\beta_2}{2},
    \end{align}
    where $\omega_1$ and $\omega_2$ are the same as in \eqref{eq:new_ensemble}, and  $W^{f \prime}_{1 \times 1}$ is defined in the same way as $W^{f}_{1 \times 1}$ in Eq.~\eqref{eq:WilsonLoop}, but excluding the links containing $U$ from the sum.
    \item Generate a new link, $U^\prime$, according to the local probability distribution $dP(U)$
    \[
        \mathrm{d} P(U) = \mathrm{d} U \exp \left(\frac{\alpha_{1}}{2} c^f_{1} \operatorname{Tr}\left[U A^f_{1}\right]+\frac{\alpha_{2}}{2} c^f_{2} \operatorname{Tr}\left[U A^f_{2}\right]\right)
    \] Again, using the property that the sum of elements of $SU(2)$ is proportional to an element of $SU(2)$
    \[
    \frac{\alpha_{1}}{2} c^f_{1} A^f_{1}+\frac{\alpha_{2}}{2} c^f_{2} A^f_{2} = \beta V,
    \] where $V\in SU(2)$ and $\beta \in \mathbb{R}$. After this manipulation, we apply the standard heatbath for $SU(2)$, see e.g.~\cite{Creutz:1980zw,Gattringer:2010zz}.
    \item The probability distribution used in 2 is not the correct one; we now need to correct for the biased quadratic term. To this end, we accept the element generated in 2 with probability
    \[
   \min \left(1, \exp \left(\omega_{2}  \cdot c^{f\,2}_1 \cdot  \frac{ \operatorname{Tr}\left[U^{\prime} A^f_{1}\right]^{2}- \operatorname{Tr}\left[U A^f_{1}\right]^{2}}{N_{\text {Loops }}}\right)\right)
    \]

\end{enumerate}
 As the steps are local, the numerator will always be much smaller than the denominator. In practical terms, for lattices of $L=4$ we obtained acceptance probabilities around $99.8\%$, allowing us to take advantage of the efficiency of the heat bath at the same time we used our modified ensemble. Note also that this algorithm works efficiently because the quadratic part is suppressed by the number of loops. This is not the case in the adjoint extension as the adjoint trace is quadratic itself in terms of the fundamental plaquette. This is why we relied on a multiple try  algorithm of  \cite{Martino_2018} in this case.

%\bibliographystyle{unsrt}
%\bibliography{ref,5d_Yang-Mills.bib}
%\newpage

\end{document}